%% file: Paper_aos.tex
\documentclass[aos,preprint]{imsart}
\pdfoutput=1
\usepackage{amsthm, amsmath, amssymb, graphicx, caption,subcaption,color}

\RequirePackage{algorithmic}
\RequirePackage[section]{algorithm}

\usepackage{xr-hyper}
\RequirePackage{hyperref}
\RequirePackage[sort, numbers]{natbib}

\input{texfiles/definitions.tex}
\begin{document}

\begin{frontmatter}

% "Title of the paper"
\title{Estimating the effect of joint interventions from observational data in sparse high-dimensional settings}
\runtitle{The effect of joint interventions from observational data}

\thankstext{T1}{Supported in part by Swiss NSF Grant 200021\_149760.}

\begin{aug}
\author{\fnms{Preetam} \snm{Nandy}\thanksref{m1,T1}\ead[label=e1]{nandy@stat.math.ethz.ch},}
%\address{\printead{e1}}
\author{\fnms{Marloes H.} \snm{Maathuis}\thanksref{m1,T1}\ead[label=e2]{maathuis@stat.math.ethz.ch}}
%\address{\printead{e2}}
\and
\author{\fnms{Thomas S.} \snm{Richardson}\thanksref{m2}\ead[label=e3]{thomasr@u.washington.edu}}

\affiliation{ETH Z\"urich\thanksmark{m1} and University of Washington\thanksmark{m2}}

\runauthor{P. Nandy, M. H. Maathuis and T. S. Richardson}

  \address{P. Nandy\\
  M. H. Maathuis\\
  ETH Zurich\\
  Seminar for Statistics\\
          R\"amistrasse 101\\
          8092 Zurich, Switzerland\\
          \printead{e1}\\
           \printead{e2}}

             \address{T. S. Richardson\\
Department of Statistics\\
              University of Washington \\
              Seattle, Washington 98195 \\
              USA\\
\printead{e3}}
 \end{aug}

\begin{abstract}
We consider the estimation of joint causal effects from observational data. In particular, we propose new methods to estimate the effect of multiple simultaneous interventions (e.g., multiple gene knockouts), under the assumption that the observational data come from an unknown linear structural equation model with independent errors. We derive asymptotic variances of our estimators when the underlying causal structure is partly known, as well as high-dimensional consistency when the causal structure is fully unknown and the joint distribution is multivariate Gaussian. We also propose a generalization of our methodology to the class of nonparanormal distributions. We evaluate the estimators in simulation studies and also illustrate them on data from the DREAM4 challenge.
\end{abstract}

%\begin{keyword}[class=MSC]
%\kwd[Primary ]{}
%\kwd{}
%\kwd[; secondary ]{}
%\end{keyword}

%\begin{keyword}
%\kwd{}
%\kwd{}
%\end{keyword}


\end{frontmatter}
%all sections
\input{texfiles/introduction}

\input{texfiles/problem_definition}

\input{texfiles/multiple_interventions_using_parents_sets}

\input{texfiles/theoretical_properties}
\input{texfiles/parents_sets_from_cpdag}

\input{texfiles/multiple_interventions_from_data}

\input{texfiles/NPN-jointIDA}

\input{texfiles/discussion}

\input{texfiles/acknowledgements}

%% AOS,AOAS: If there are supplements please fill:
%\begin{supplement}[id=supplement]
%%  \sname{Supplement A}
%  \stitle{Supplement to ``Estimating the effect of joint interventions from observational data in sparse high-dimensional settings"}
%%  \slink[doi]{10.1214/00-AOASXXXXSUPP}
% \sdatatype{supplement.pdf}"
% \sdescription{All proofs, simulation results, and an illustration of our methods to the DREAM4 challenge \citep{MarbachEtAl10} can be found in the supplementary material \citep{NandyMaathuisRichardson14b}.}
%\end{supplement}

% the bibliography
\bibliographystyle{imsart-number}
\bibliography{bib/Mybibliography}

\end{document}

%% file: texfiles/definitions.tex
\newtheorem{theorem}{Theorem}[section]
\newtheorem{corollary}{Corollary}[section]
\newtheorem{remark}{Remark}[section]
\newtheorem{definition}{Definition}[section]
\newtheorem{example}{Example}
\newtheorem*{assumption*}{Assumption}

\newcommand{\Prob}{\mathbb{P}}

\newcommand{\Exp}{\mathrm{E}}
\newcommand{\Cov}{\mathrm{Cov}}
\newcommand{\Var}{\mathrm{Var}}
\newcommand{\Dag}{\mathcal{G}}

\newcommand{\st}{:}

\newcommand{\allvar}{\mathbf{X}}
\newcommand{\rvX}{X}
\newcommand{\rvx}{x}

\newcommand{\rvY}{\rvX_p}

\newcommand{\VSet}{\mathbf{V}}
\newcommand{\ESet}{\mathbf{E}}
\newcommand{\WMat}{B}

\newcommand{\PaVec}{\mathbf{PA}}

\newcommand{\NdVec}{\mathbf{ND}}

\newcommand{\NeSet}{\mathbf{ADJ}}
\newcommand{\effectvec}{\boldsymbol\theta_{p}}
\newcommand{\effectvecn}{\boldsymbol\theta_{p_n}}
\newcommand{\effect}{\theta}

\newcommand{\rvW}{\mathbf{W}}

\newcommand{\errorvec}{\boldsymbol\epsilon}
\newcommand{\error}{\epsilon}

\newcommand{\density}{f}

\newcommand{\impvar}{\mathbf{U}}

\newcommand{\covmat}{\Sigma}
\newcommand{\sampcovmat}{\hat{\covmat}}

\newcommand{\chol}{L}

\newcommand{\diagmat}{D}

\newcommand{\halfvec}{\mathrm{vech}}

%% file: texfiles/introduction.tex
\section{Introduction}\label{section: introduction}
Estimation of causal effects from observational data is impossible in general. It is, however, possible to estimate sets of possible causal effects of single interventions from observational data, under the assumption that the data were generated from an unknown linear structural equation model (SEM) with independent errors, or equivalently, from an unknown directed acyclic graph (DAG) model without hidden variables. The IDA method \citep{MaathuisKalischBuehlmann09} was developed for this purpose, and can for example be used to predict the effect of single gene knockouts on other genes or some phenotype of interest, based on observational gene expression profiles. IDA has been applied to high-dimensional gene expression data sets and is a useful tool for the design of experiments, in the sense that it can indicate which genes are likely to have a large effect on the variable of interest \citep{MaathuisColomboKalischBuehlmann10, StekhovenEtAl12}.

In this paper, we generalize IDA by relaxing some of its assumptions and extending it to \emph{multiple} simultaneous interventions. For example, we may want to predict the effect of a double or triple gene knockout on some phenotype of interest. Since the space of possible intervention experiments grows exponentially in the number of simultaneous interventions, having an IDA-like tool to predict the effect of multiple simultaneous interventions is highly desirable in order to plan and prioritize such experiments. Moreover, the strength of the \emph{epistatic interaction} \citep{JasnosKorona07, VelenichGore13} between a pair of genes can be estimated by computing the difference between the predicted effect of a double gene knockout and the combined predicted effects of the two corresponding single gene knockouts.

The idea behind IDA is as follows. Since the underlying causal DAG is unknown, it seems natural to try to learn it. In general, however, the underlying causal DAG is not identifiable. We can learn its Markov equivalence class, which can be represented as a graph by a so-called \emph{completed partially directed acyclic graph} (CPDAG) (see Section \ref{A2-section: Markov equivalence class of DAGs} of \citep{NandyMaathuisRichardson14b}). Conceptually, we can then list all DAGs in the Markov equivalence class. One of these DAGs is the true causal DAG, but we do not know which one. For each DAG, we can then estimate the total causal effect of say $X_i$ on $X_p$, under the assumption that the given DAG is the true causal DAG. In a linear SEM, this means that we can simply take the coefficient of $X_i$ in the regression of $X_p$ on $X_i$ and the parents of $X_i$ in the given DAG (see Section \ref{subsection: covariate adjustment}). Doing this for all DAGs in the Markov equivalence class yields a multiset of possible causal effects that is guaranteed to contain the true causal effect. We can then summarize the information on the effect of $X_i$ on $X_p$ by taking summary measures of this multiset.

For large graphs, listing all DAGs in the Markov equivalence class is computationally intensive. The above reasoning shows, however, that it suffices to know the parents of $X_i$ in the different DAGs. These possible parent sets can be extracted directly from the CPDAG, using a simple local criterion \citep{MaathuisKalischBuehlmann09}. This approach has two important advantages: it is a computational shortcut and it is less sensitive to estimation errors in the estimated CPDAG. The three steps of IDA can then be summarized as
(1) estimating the CPDAG; (2) extracting possible valid parent sets of the intervention node $X_i$ from the CPDAG; (3) regressing $X_p$ on $X_i$ while adjusting for the possible parent sets.
A schematic representation of IDA is given in Section \ref{A2-subsection: IDA for single intervention} of \citep{NandyMaathuisRichardson14b}.

In order to generalize IDA to estimate the effect of joint interventions, we need non-trivial modifications of steps (2) and (3).
In step (2), we must make sure that the possible parent sets of the various intervention nodes are jointly valid, in the sense that there is a DAG in the Markov equivalence class with this specific combination of parent sets. This decision can no longer be made fully locally, as was possible for the single intervention case.
In step (3), we can no longer use regression with covariate adjustment, as illustrated in Example \ref{ex: path method 2} in Section \ref{subsection: covariate adjustment} (cf. \citep{ShpitserVanderWeeleRobins10}). We therefore develop new methods to estimate the effect of joint interventions under the assumption that only the parent sets of the intervention nodes are given. We refer to this assumption as the OPIN assumption (\underline{o}nly \underline{p}arents of \underline{i}ntervention \underline{n}odes).

In the literature on time-dependent treatments (which can be viewed as joint interventions), it has been proposed to use inverse probability weighting (IPW) \citep{RobinsHernanBrumback00}. IPW fits our framework in the sense that it works under the OPIN assumption. The method is widely used when the underlying causal DAG is given, but combining it with a causal structure learning method seems new. Unfortunately, however, such a combination does not provide a satisfactory solution to our problem, since we found that the statistical behavior was disappointing in our setting with continuous treatments. We therefore propose two new methods for estimating the effect of joint interventions under the OPIN assumption: one is based on \underline{r}ecursive \underline{r}egressions for \underline{c}ausal effects (RRC) and the other on \underline{m}odifying \underline{C}holesky \underline{d}ecompositions (MCD). Combining our new steps (2) and (3), we obtain methods, called \emph{jointIDA}, for estimating the effect of joint interventions from observational data, under the assumption that the data were generated from an unknown linear SEM with continuous independent errors. 

We establish asymptotic normality of our estimators when the underlying SEM is linear and the parent sets are known (Section \ref{section: theoretical properties}). Moreover, we provide high-dimensional consistency results when the causal structure is fully unknown and the distribution is multivariate Gaussian (Section \ref{section: estimation from observational data}). We also provide a generalization of our methodology to the family of nonparanormal distributions (Section \ref{section: relaxing linearity}).

Compared to the original IDA method \cite{MaathuisKalischBuehlmann09}, we have considerably weaker assumptions. IDA required linearity, Gaussianity and no hidden confounders. We dropped the Gaussianity assumption to a large extent, and only use it now in the high-dimensional consistency proof of (joint-)IDA, where it is needed since (so far) no causal structure learning method has been shown to be consistent in high-dimensional settings for linear SEMs with non-Gaussian noise. Additionally, we give an extension of our methods to nonparanormal distributions, hence treating an interesting subclass of non-linear and non-Gaussian distributions.

%\textcolor{red}{IDA was developed under the assumption of Gaussian errors, while we only assume continuous distributions of the error variables. To this end, we combine RRC and MCD with the Rank-PC algorithm of \citep{HarrisDrton13} which has been shown to be more robust under non-Gaussian errors than the PC algorithm \citep{SpirtesEtAl00} that was used in IDA for estimating the underlying CPDAG. }

%This paper is organized as follows. Section \ref{section: problem definition} defines the problem and discusses necessary background knowledge. Section \ref{section: multiple interventions using parent sets} considers estimation of the total joint effect of multiple simultaneous interventions under the OPIN assumption, and introduces our new RRC and MCD estimators. Section \ref{section: theoretical properties} gives the corresponding asymptotic distributions. Section \ref{section: parent sets from cpdag} describes methods for extracting jointly valid parent sets for multiple intervention nodes from a CPDAG. Section \ref{section: estimation from observational data} combines the methods in Sections 3 and 5 into a joint-IDA estimator that can be applied to observational data from an unknown linear SEM. This section also includes consistency results for sparse high-dimensional settings under the assumption of Gaussian errors. Section \ref{section: application} contains an application to data from the DREAM4 challenge \citep{MarbachEtAl09}, followed by a discussion in Section \ref{section: discussion}. 
All proofs, simulation results, and an illustration of our methods to the DREAM4 challenge \citep{MarbachEtAl10} can be found in the supplementary material \citep{NandyMaathuisRichardson14b}. Joint-IDA has been implemented in the \texttt{R}-package \textbf{pcalg} \citep{KalischEtAl12}.

%% file: texfiles/problem_definition.tex
\section{Preliminaries}\label{section: problem definition}

%We now define the problem precisely. We assume that the data are generated by an unknown linear structural equation model with independent additive errors, as specified in Section \ref{subsection: linear SEM}. Our goal is to estimate the effect of multiple simultaneous interventions, as defined in Section \ref{subsection: effect of multiple interventions}. The slightly more general mechanism changes are defined in Section \ref{subsection: mechanism change}.

\subsection{Graph terminology}

We consider graphs $\mathcal H = (\VSet,\ESet)$ with vertex (or node) set $\VSet$ and edge set $\ESet$. There is at most one edge between any pair of vertices and edges may be either directed ($i\to j$) or undirected ($i - j$). If $\mathcal H$ contains only (un)directed edges, it is called \emph{(un)directed}. If $\mathcal H$ contains directed and/or undirected edges, it is called \emph{partially directed}. The \emph{skeleton} of a partially directed graph is the undirected graph that results from replacing all directed edges by undirected edges.

If there is an edge between $i$ and $j$ in $\mathcal H$, we say that $i$ and $j$ are \emph{adjacent}. The adjacency set of $i$ in $\mathcal H$ is denoted by $\NeSet_i(\mathcal H)$. If $i \to j$ in $\mathcal H$, then $i$ is a \emph{parent} of $j$, and the edge between $i$ and $j$ is \emph{into} $j$. The set of all parents of $j$ in $\mathcal H$ is denoted by $\PaVec_j(\mathcal H)$.

A \emph{path between $i$ and $j$} is a sequence of distinct vertices $(i,\dots,j)$ such that all pairs of successive vertices are adjacent. A \emph{backdoor path} from $i$ to $j$ is a path between $i$ and $j$ that has an edge into $i$, i.e., $i \leftarrow \dots \dots j$. A path $(i,j,k)$ is called a \emph{v-structure} if $i\to j \leftarrow k$ and $i$ and $k$ are not adjacent. A \emph{directed path from $i$ to $j$} is a path between $i$ and $j$ where all edges are directed towards $j$. A directed path from $i$ to $j$ together with the edge $j \to i$ forms a \emph{directed cycle}. If there is a directed path from $i$ to $j$, then $i$ is an \emph{ancestor} of $j$ and $j$ is a \emph{descendant} of $i$. We also say that each node is an ancestor and a descendant of itself. The set of all non-descendants of $i$ in $\mathcal H$ is denoted by $\NdVec_i(\mathcal H)$.

A graph that does not contain directed cycles is called \emph{acyclic}. Important classes of graphs in this paper are \emph{directed acyclic graphs (DAGs)} and \emph{partially directed acyclic graphs (PDAGs)}.

%%%%%%%%%%%%%%%%%%%%%%%%%%%%%%%%%%%%%%%%%%%%%%%%%%%%%%%%%%%%%%%%%%%%%%%%%%%%%%%%%
\subsection{Linear structural equation models and causal effects}\label{subsection: linear SEM}

Throughout this paper, we use the same notation to refer to sets or vectors. For example, $\allvar$, $\NeSet$ and $\PaVec$ can refer to sets or vectors, depending on the context.

Let $(\VSet, \ESet)$ be a DAG with $|\VSet|=p$ vertices. Each vertex $i \in \VSet$, $i\in \{1,\dots,p\}$, represents a random variable $\rvX_i$. An edge $i\to j$ means that $X_i$ is a direct cause of $X_j$ in the sense of Definition \ref{definition: linear SEM} below. Throughout this paper, we use the same notation to denote a set of vertices and the corresponding set of random variables. For example, $\PaVec_i(\mathcal{H})$ can refer to a set of indices or a set of random variables. Let $\WMat$ be a $p \times p$ weight matrix, where $\WMat_{ij}$ is the weight of the edge $i\to j$ if $i\to j \in  \ESet$, and $\WMat_{ij} = 0$ otherwise. Then we say that $\Dag = (\VSet,\ESet,\WMat)$ is a weighted DAG.

\begin{definition}\label{definition: linear SEM} (Linear structural equation model)
   Let $\Dag = (\VSet,\ESet,\WMat)$ be a weighted DAG, and $\errorvec = (\error_1,\ldots,\error_p)^{T}$ a continuous random vector of jointly independent error variables with mean zero. Then $\allvar = (\rvX_1,\dots,\rvX_p)^T$ is said to be generated from a linear structural equation model (linear SEM) characterized by the pair $ (\Dag,\errorvec)$ if% $\allvar$ satisfies
   \vspace{-0.1in}
   \begin{align}\label{eq: linear SEM 1}
      \allvar \leftarrow \WMat^T \allvar + \errorvec.
   \end{align}
\end{definition}

If $\allvar$ is generated from a linear SEM characterized by the pair $ (\Dag,\errorvec)$ with $\Dag=(\VSet,\ESet,\WMat)$, then we call $\Dag$ the \emph{causal weighted DAG} and  $(\VSet,\ESet)$ the \emph{causal DAG}. The symbol ``$\leftarrow$" in (\ref{eq: linear SEM 1}) emphasizes that the expression should be understood as a generating mechanism rather than as a mere equation.

We emphasize that we assume that there are no hidden variables; hence the joint independence of the error terms. In the rest of the paper, we refer to linear SEMs without explicitly mentioning the independent error assumption. We also consider each of the $p$ structural equations in \eqref{eq: linear SEM 1} as ``autonomous", meaning that changing the generating mechanism of one of the variables does not affect the generating mechanisms of the other variables.
%If all error variables are Gaussian, then $\allvar$ is multivariate Gaussian and we refer to \eqref{definition: linear SEM} as a \emph{Gaussian linear SEM}.

An example of a weighted DAG $\Dag$ is given in Figure \ref{fig: path method}, where $p=6$, $\rvX_1 \leftarrow 0.2 \rvX_5 + \epsilon_1$, $\rvX_2 \leftarrow 0.6 \rvX_3 +  0.5\rvX_4 + \epsilon_2$, $\rvX_3 \leftarrow 1.1 \rvX_1 + \epsilon_3$, $\rvX_4 \leftarrow 0.3\rvX_1 + 0.8\rvX_3 + 0.7\rvX_5 + \epsilon_4$, $\rvX_5 \leftarrow \epsilon_5$ and $\rvX_6 \leftarrow 0.4\rvX_2 + 0.9\rvX_3 + \epsilon_6$.
%The weight matrix $B$ is given in Section \ref{A2-subsection: an illustration of the MCD algorithm} of \cite{NandyMaathuisRichardson14b}.
Note that $X_5$ directly causes $X_1$, in the sense that $X_5$ plays a role in the generating process of $X_1$. The set of all direct causes of $\rvX_i$ is $\PaVec_i(\Dag)$.

Suppose that $\allvar$ is generated from a linear SEM characterized by $(\Dag, \errorvec)$. Since $\Dag$ is acyclic, the vertices can always be rearranged to obtain an upper triangular weight matrix B. Such an ordering of the nodes is called a \emph{causal ordering}. In Figure \ref{fig: path method}, (5, 1, 3, 4, 2, 6) is a causal ordering. (In this example there is a unique causal ordering, but that is not true in general.)

For any $\allvar$ generated by a linear SEM characterized by $(\Dag, \errorvec)$, the joint density of $\allvar$ satisfies the following factorization \cite{Pearl09}:
\vspace{-0.1in}
\begin{equation*}
   \density(\rvx_1,\ldots,\rvx_p) = \prod_{i=1}^{p} \density (\rvx_i | \mathbf{pa}_i),\vspace{-0.1in}
\end{equation*}
where the parent sets $\mathbf{pa}_i = \mathbf{pa}_i(\Dag)$ are determined from $\Dag$.

We now consider a (hypothetical) outside intervention to the system, where we set a variable $X_j$ to some value $x_j'$ within the support of $X_j$, uniformly over the entire population. This can be denoted by Pearl's do-operator: $do(X_j = x_j')$ or $do(x_j')$ \cite{Pearl09}, which corresponds to removing the edges into $X_j$ in $\Dag$ (or equivalently, setting the $j$th column of $B$ equal to zero) and replacing $\error_j$ by the constant $x_j'$. Since we assume that the other generating mechanisms are not affected by this intervention, the post-intervention joint density is given by the so-called truncated factorization formula or g-formula, see \cite{Pearl09, Robins86, SpirtesEtAl00}:
%, since the generating mechanism of $X_j$ no longer depends on its (former) parents.
\begin{align*}
   \density(\rvx_1,\ldots,\rvx_p | do(\rvX_j = \rvx_j^{\prime})) = \left\{ \begin{array}{ll} \prod_{i\neq j} \density(\rvx_i | \mathbf{pa}_i) & \text{if $\rvx_j = \rvx_j^{\prime}$ } \\ 0 & \text{otherwise} \end{array}  \right..
\end{align*}
%where the term $\density(\rvx_j | \mathbf{pa_j})$ is omitted from the factorization.

The post-intervention distribution after a joint intervention on several nodes can be handled similarly:
\begin{align}\label{eq: truncated factorization}
\density(\rvx_1,\ldots,\rvx_p | do(\rvx_1^{\prime},\ldots,\rvx_k^{\prime})) = \left\{ \begin{array}{ll} \prod_{i>k} \density(\rvx_i | \mathbf{pa}_i) & \text{if $\rvx_i = \rvx_i^{\prime}$ $\forall ~ i \leq k$} \\ 0 & \text{otherwise} \end{array}  \right..
\end{align}

Unless stated otherwise, we assume that $(\rvX_1,\dots,\rvX_k)$ are the intervention variables ($k \in \{1,\dots,p-1\}$) and $\rvY$ is the variable of interest. One can always label the variables to achieve this, since the nodes are not assumed to be in a causal ordering. The number of intervention variables is called the \emph{cardinality} of the joint intervention.

Definition \ref{definition: total causal effect} defines the total joint effect of $(X_1,\dots,X_k)$ on $\rvY$ in terms of partial derivatives of the expected value of the post-intervention distribution of $\rvY$.
\begin{definition}\label{definition: total causal effect} (Total joint effect)
   Let $\allvar$ be generated from a linear SEM characterized by $(\Dag,\errorvec)$. Then the total joint effect of $(X_1,\dots,X_k)$ on $\rvY$ is given by $$\effectvec^{(1,\ldots,k)} := (\effect_{1p}^{(1,\ldots,k)},\ldots,\effect_{kp}^{(1,\ldots,k)})^T,$$
   where \vspace{-0.1in}$$\effect_{ip}^{(1,\ldots,k)} := \frac{\partial}{\partial \rvx_i} \Exp[\rvY | do(\rvx_1,\ldots,\rvx_k)], \quad \text{for}~ i = 1,\ldots,k,$$
   is the total effect of $\rvX_i$ on $\rvY$ in a joint intervention on $(\rvX_1,\dots,\rvX_k)$.
   For notational convenience, we write $\effect_{ip}$ instead of $\effect_{ip}^{(i)}$ to denote the total effect of $\rvX_i$ on $\rvY$ in a single intervention on $\rvX_i$. Finally, we write $\effectvec^{(1,\dots,k)}(\Dag)$ and $\effect_{ip}^{(1,\dots,k)}(\Dag)$ when it is helpful to indicate the dependence on the weighted DAG $\Dag$.
\end{definition}
In general, $\effectvec^{(1,\ldots,k)}$ is a vector of functions of $x_1,\dots,x_k$, but under the assumption that $\allvar$ is generated from a linear SEM, it reduces to a vector of numbers. In this case the partial derivatives can be interpreted as follows:
\begin{align*}
 \effect_{ip}^{(1,\ldots,k)} = \Exp[\rvY | do(\rvx_1,\ldots,\rvx_i+1,\ldots, \rvx_k)]  - \Exp[\rvY | do(\rvx_1,\ldots,\rvx_i,\ldots, \rvx_k)].
\end{align*}
Thus, the total effect of $\rvX_i$ on $\rvX_p$ in a joint intervention on $(\rvX_1,\ldots,\rvX_k)$ represents the increase in expected value of $\rvY$ due to one unit increase in the intervention value of $\rvX_i$, while keeping the intervention values of $\{\rvX_1,\ldots,\rvX_k\}\setminus\{\rvX_i\}$ constant. (In certain cases, $\effect_{ip}^{(1,\dots,k)}$ can be viewed as a \emph{direct} effect; see for example $\theta_{1p}^{(1,2)}$ in Figure \ref{A2-subfig: RRCvsCochran1} in Section \ref{A2-subsection: Cochran} of \cite{NandyMaathuisRichardson14b}.)

The meaning of $\effect_{ip}^{(1,\ldots,k)}$ in a linear SEM can also be understood by looking at the causal weighted DAG $\Dag$: the causal effect of $\rvX_i$ on $\rvX_p$ along a directed path from $i$ to $j$ in $\Dag$ can be calculated by multiplying all edge weights along the path, see \cite{Wright21}.  Then each $\effect_{ip}^{(1,\ldots,k)}$ can be calculated by summing up the causal effects along all directed paths from $i$ to $p$ which do not pass through $\{1,\ldots,k\}\setminus \{i\}$ (since those variables are held fixed by the intervention). We refer to this interpretation as the ``path method" and illustrate it in the following example.
\begin{example}\label{ex: path method}
\vspace{-0.15in}
    \begin{figure}[!ht]
       \centering
       \includegraphics[width=0.3\textwidth]{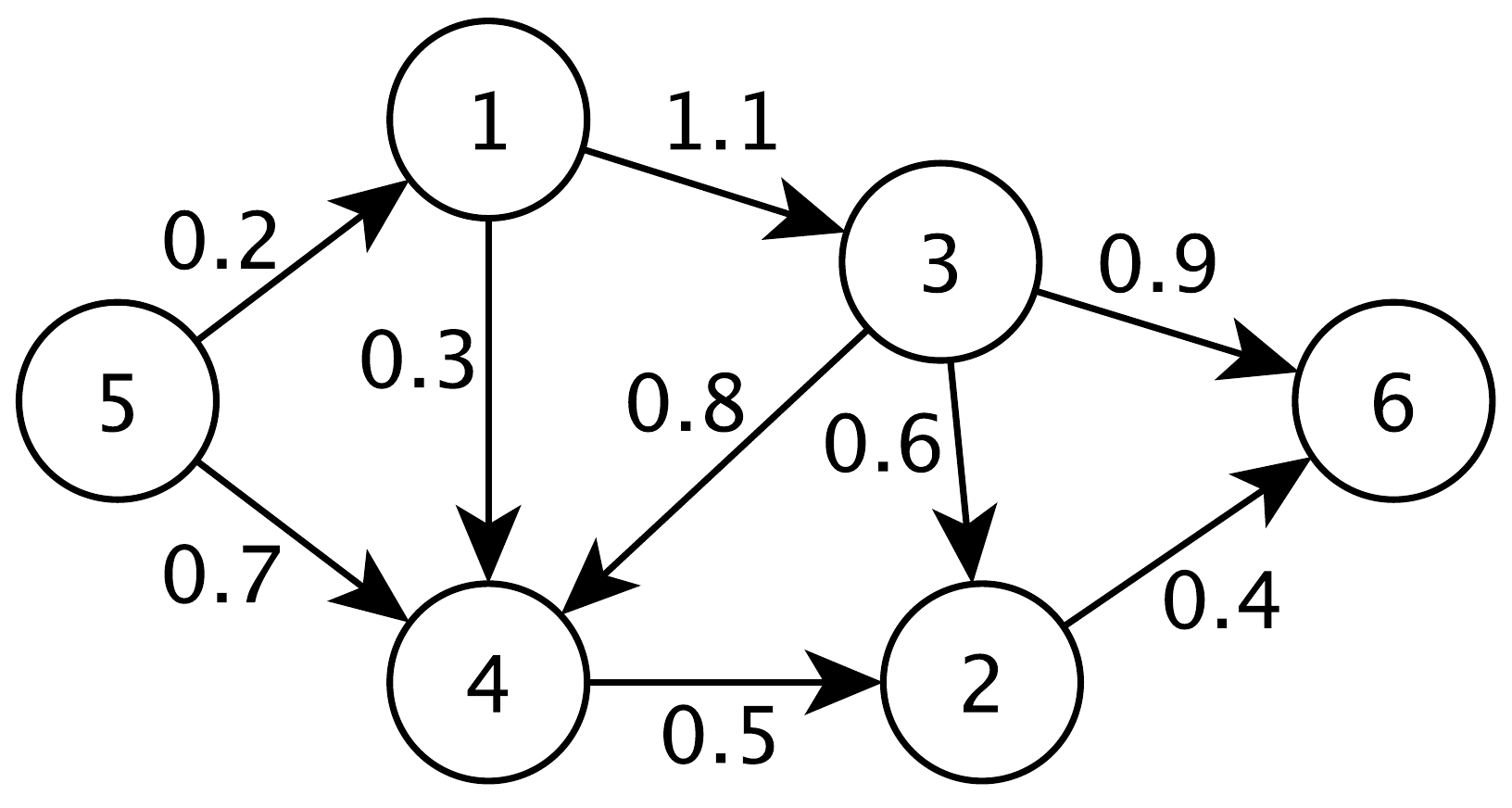}
       \caption{A weighted causal DAG $\Dag$.}
       %\caption{The weighted DAG $\Dag$ which represents the causal structure of the variables $\rvX_1,\ldots,\rvX_5$ in Example \ref{ex: path method}.}
       \label{fig: path method}
          \vspace{-0.1in}
    \end{figure}
    Let $\rvX_1,\ldots,\rvX_6$ be generated from a linear SEM characterized by $(\Dag,\errorvec)$, where $\Dag = (\VSet,\ESet,\WMat)$ is depicted in Figure \ref{fig: path method} and $\errorvec=(\epsilon_1,\dots,\epsilon_6)^T$ are jointly independent errors with arbitrary mean zero distributions.

    We first consider the total effect of $\rvX_1$ on $\rvX_6$ in a single intervention on $\rvX_1$. There are four directed paths from $1$ to $6$, namely $1\rightarrow 3 \rightarrow 6$, $1\rightarrow 3 \rightarrow 2 \rightarrow 6$, $1 \rightarrow 3 \rightarrow 4 \rightarrow 2 \rightarrow 6$ and $1 \rightarrow 4 \rightarrow 2 \rightarrow 6$.  Hence, the total causal effect of $\rvX_1$ on $\rvX_6$ is $\effect_{16} = \WMat_{13}\WMat_{36} + \WMat_{13}\WMat_{32}\WMat_{26} + \WMat_{13} \WMat_{34} \WMat_{42}\WMat_{26}+\WMat_{14}\WMat_{42}\WMat_{26}= 1.1 \times 0.9 + 1.1\times0.6\times 0.4 + 1.1\times0.8\times 0.5\times 0.4 + 0.3\times 0.5 \times 0.4= 1.49$. Similarly, the total causal effect of $\rvX_2$ on $\rvX_6$ is $\effect_{26} = \WMat_{26} = 0.4$.

 Next, we consider the total joint effect of $(\rvX_1,\rvX_2)$ on $\rvX_6$. Since the only directed path from $2$ to $6$ $(2\to6)$ does not pass through $1$, $\effect_{26}^{(1,2)}= \effect_{26}$. On the other hand, three of the four directed paths from 1 to 6 pass through 2, and the only remaining directed path is $1\rightarrow 3 \rightarrow 6$. Hence, $\effect_{16}^{(1,2)} =  \WMat_{13}\WMat_{36} = 1.1\times 0.9 = 0.99$, which is different from the single intervention effect $\effect_{16}=1.49$.
\end{example}

\begin{remark}\label{remark: thetaij=0}
    If $X_j \in \NdVec_i(\Dag)$, then there is no directed path from $i$ to $j$ in $\Dag$. Thus, $\theta_{ij}=0$ and the total effect of $\rvX_i$ on $\rvX_j$ is also zero in any joint intervention that involves $\rvX_i$.

  %Thus, $\theta_{ij}$ is zero. Since $\Dag$ is acyclic, at most one of $\theta_{ij}$ and $\theta_{ji}$ can be nonzero.
%  Moreover, it follows directly from the path method that if $\theta_{ij}$ is zero, then there is no directed path from $i$ to $j$ in the weighted \textcolor{red}{causal} DAG $\Dag$ (except for some very special choices of the edge weights), and hence the total effect of $\rvX_i$ on $\rvX_j$ is also zero in any joint intervention that involves $\rvX_i$.
\end{remark}

%%%%%%%%%%%%%%%%%%%%%%%%%%%%%%%%%%%%%%%%%%%%%%%%%%%%%%%%%%%%%%%%%%%%%%%%%%%%%%%%%
\subsection{Causal effects via covariate adjustment}\label{subsection: covariate adjustment}

It is straightforward to determine the total effect of $\rvX_1$ on $\rvY$ in a single intervention on a linear SEM (see Proposition \ref{A2-proposition: causal effect in linear SEM} of \cite{NandyMaathuisRichardson14b}), since\vspace{-0.05in}
\begin{align}\label{eq: causal effect}
  \effect_{1p} =   \left\{ \begin{array}{ll} 0 & \text{if $\rvY \in \PaVec_1$} \\  \beta_{1p | \PaVec_1} & \text{otherwise} \end{array} \right.,\vspace{-0.05in}
\end{align}
where, for any $j \neq i$ and any set of variables $\mathbf{S}$ such that $\{\rvX_i,\rvX_j\}\cap \mathbf{S} = \emptyset$, we define $\beta_{ij | \mathbf{S}} $ to be the coefficient of $\rvX_i$ in the linear regression of $\rvX_j$ on $\{\rvX_i\} \cup \mathbf{S}$ (without intercept term), denoted by $\rvX_j \sim \rvX_i + \mathbf{S}$ \cite{MaathuisKalischBuehlmann09}. Equation \eqref{eq: causal effect} immediately follows from Pearl's backdoor criterion \cite{Pearl09} if all error variables are Gaussian \cite{MaathuisKalischBuehlmann09}. In Section \ref{A2-section: single intervention effects as regression coefficients} of \cite{NandyMaathuisRichardson14b}, we show that \eqref{eq: causal effect} in fact holds under a linear SEM with arbitrary error distributions, since both the left hand side and the right hand side only depend on the covariance matrix of $\allvar$.
%\textcolor{red}{if all error variables are Gaussian}. \textcolor{red}{However, without the Gaussian assumption the result is not straightforward, mainly because $\Exp[\rvX_p | \rvx_1, \paVec_1]$ may then be a nonlinear function of $\rvx_1$ and $\paVec_1$. We present a proof of \eqref{eq: causal effect} in Section \ref{A2-section: single intervention effects as regression coefficients} of \cite{NandyMaathuisRichardson14b}.}
%\begin{remark}\label{remark: causal effects via unadjusted regressions}
%If there is no path from $\rvX_1$ to $\rvX_p$ that starts with an edge into $\rvX_1$, then $\effect_{1p} = \beta_{1p | \emptyset}$. This follows from the fact that the empty set satisfies Pearl's backdoor criterion in this case (see Remark \ref{A2-remark: backdoor set} of \cite{NandyMaathuisRichardson14b}).
%\end{remark}
Comparing equation \eqref{eq: causal effect} to the path method, we see that \eqref{eq: causal effect} does not require any knowledge about the underlying causal DAG beyond the parents of the intervention node $\rvX_1$.

%Moreover, we note that $\effect_{1p}=0$ if $\rvY$ is a non-descendant of $\rvX_1$: if $\rvY \in \PaVec_1$ then $\effect_{1p}=0$ by definition, and if $\rvY$ is a non-descendant but not a parent of $\rvX_1$ then $\beta_{1p | \PaVec_1} = 0$.

For the total joint effect of $(\rvX_1,\dots,\rvX_k)$ on $\rvY$ ($k>1$), straightforward covariate adjustment cannot be used to calculate $\effectvec^{(1,\ldots,k)}$ from one multiple linear regression. One might perhaps hope that each $\effect_{ip}^{(1,\ldots,k)}$ ($i=1,\dots,k$) can be computed separately as a coefficient of $\rvX_i$ in a multiple linear regression, but the following example shows that this strategy fails as well.

\begin{example}\label{ex: path method 2}
 We reconsider Example \ref{ex: path method} with the causal weighted DAG $\Dag$ in Figure \ref{fig: path method}. %Let $\allvar = \{\rvX_1,\ldots,\rvX_6\}$ be generated from the linear SEM characterized by $(\Dag,\errorvec)$, where $\Cov(\errorvec)$ is a diagonal matrix.
Then $\effect_{16}^{(1,2)}$ cannot be computed as the coefficient of $\rvX_1$ in a multiple linear regression. This can be verified by computing the coefficients of $\rvX_1$ in all $2^4$ regressions $\rvX_6 \sim \rvX_1 + \mathbf{S}$ for $\mathbf{S} \subseteq \{\rvX_2,\rvX_3,\rvX_4,\rvX_5\}$. None of these coefficients equal $\effect_{16}^{(1,2)}=0.99$ as obtained from the path method.

\end{example}

%\subsection{Markov equivalence class of DAGs}\label{subsection: Markov equivalence class of DAGs}
%
% A DAG encodes conditional independence relationships via the notion of \textit{d-separation} (\cite{Pearl00}, Theorem 1.2.4, page 18). In general, several DAGs can encode the same conditional independence relationships and such DAGs form a \emph{Markov equivalence class}. Two DAGs belong to the same Markov equivalence class if and only if they have the same skeleton and the same v-structures \cite{VermaPearl90}. A Markov equivalence class of DAGs can be uniquely represented by a \emph{completed partially directed acyclic graph} (CPDAG) \cite{Chickering02}, which is a graph that can contain both directed and undirected edges. Figure \ref{fig: a simple CPDAG} shows an example of a CPDAG, as well as the DAGs in its Markov equivalence class. A CPDAG satisfies the following: $i\to j$ in the CPDAG if $i\to j$ in every DAG in the Markov equivalence class, and $i - j$ in the CPDAG if the Markov equivalence class contains a DAG for which $i\to j$ as well as a DAG for which $i\leftarrow j$. CPDAGs can be estimated from observational data using various algorithms \cite{SpirtesEtAl00, Chickering03, TsamardinosEtAl06}. The PC-algorithm \cite{SpirtesEtAl00} is one such algorithm, and is consistent in sparse high-dimensional settings \cite{KalischBuehlmann07a}.

%% file: texfiles/multiple_interventions_using_parents_sets.tex
\section{Joint interventions when we only know the parents of the intervention nodes}\label{section: multiple interventions using parent sets}

Let $\allvar$ be generated from a linear SEM characterized by $(\Dag,\errorvec)$, and suppose that we are interested in the total joint effect of $(\rvX_1,\dots,\rvX_k)$ on $\rvY$. If $\Dag$ were known, then these effects could be computed with the path method. In this section, we consider the following weaker assumption.
\vspace{-0.1in}
\begin{assumption*} (OPIN: only parents of intervention nodes)
  We only have partial knowledge of the underlying DAG $\Dag$: we know the direct causes (parent sets) of the intervention variables $\rvX_1,\dots,\rvX_k$, but have no other information about the underlying causal structure. In particular, we do not know, in general, whether $i$ comes before or after $j$ in a causal ordering of the nodes for any $i\neq j$ in $\{1,\dots,k\}\cup\{p\}$.
\end{assumption*}
\vspace{-0.1in}
We consider this set-up for two main reasons. First, we think it is an interesting and novel assumption in itself, as there may be scenarios where one does not know the entire causal DAG, but one does know the direct causes of the intervention nodes. Second, it is a stepping stone for determining possible total joint effects in settings where the underlying causal DAG is fully unknown.
 %In such settings, the underlying causal DAG is generally not identifiable, but we can identify its CPDAG, representing its Markov equivalence class (see Section \ref{A2-section: Markov equivalence class of DAGs} of \cite{NandyMaathuisRichardson14b}). Conceptually, one could then list all DAGs in the Markov equivalence class and apply the path method to each of these DAGs to obtain a multiset of possible total joint effects. But this procedure quickly becomes infeasible for large graphs.
 In particular, we can mimic the IDA approach and use the CPDAG to determine possible jointly valid parent sets, i.e., parent sets of the intervention nodes that correspond to a DAG in the Markov equivalence class (see Section \ref{section: parent sets from cpdag}). For each of these possible jointly valid parent sets, we can compute the total joint effect under the OPIN assumption, and then collect all of these in a multiset. For very large graphs, one could even go a step further and only learn the Markov blankets of the intervention nodes  \cite{AliferisEtAl10, Castelo06, Ramsey06} and extract possible parent sets from there.

We say that a procedure is an \emph{OPIN method} if it does not require any knowledge of the underlying causal DAG beyond the parent sets of the intervention nodes.
As mentioned in Section \ref{section: introduction}, an existing OPIN method for joint interventions is given by IPW \cite{RobinsHernanBrumback00}.
We introduce two new OPIN methods, called RRC and MCD. Sections \ref{subsection: RRC} and \ref{subsection: MCD} discuss the ``oracle versions" of the methods, where we assume that the true distribution of $\allvar$ is fully known. The corresponding sample versions are given in Section \ref{subsection: sample versions}.

%%%%%%%%%%%%%%%%%%%%%%%%%%%%%%%%%%%%%%%%%%%%%%%%%%%%%%%%%%%%%%%%%%%%%%%%%%%%%%%%%%%%%%%%%%%%%%%%%%%%%%%%%%%%%%%%%%%%%%%%%%%%%%

\subsection{Recursive regressions for causal effects (RRC)}\label{subsection: RRC}

Our first method is based on recursive regressions for causal effects (RRC). We start with the special case of double interventions, i.e., $k=2$.
\begin{theorem}\label{theorem: soundness of RRC}
   (Oracle version of RRC for $k=2$) Let $\allvar$ be generated from a linear SEM. Then the total joint effect of $(\rvX_1,\rvX_2)$ on $\rvY$ is given by\vspace{-0.05in}
   \begin{align}\label{eq: RRC double intervention}
     \effectvec^{(1,2)} = (\effect_{1p}^{(1,2)},\effect_{2p}^{(1,2)})^T = (\effect_{1p} - \effect_{12} \effect_{2p},~ \effect_{2p} - \effect_{21} \effect_{1p})^T,\vspace{-0.05in}
   \end{align}
   where $\theta_{ij}$ is defined in \eqref{eq: causal effect}.
\end{theorem}
This result may seem rather straightforward, but we were unable to find it in the literature. There is a somewhat similar recursive formula for regression coefficients \cite{Cochran38}, namely $\beta_{1p | 2} = \beta_{1p} - \beta_{12}\beta_{2p | 1} $, which is considered in the causality context (e.g., \cite{CoxWermuth03, CoxWermuth04}). However, the expression for $\effect_{1p}^{(1,2)}$ in \eqref{eq: RRC double intervention} contains causal effects which are generally different from the corresponding regression coefficients in $\beta_{1p | 2} = \beta_{1p} - \beta_{12}\beta_{2p | 1} $ (see equation \eqref{eq: causal effect} above and Example \ref{A2-ex: RRC vs Cochran's formula} in Section \ref{A2-subsection: Cochran} of \cite{NandyMaathuisRichardson14b}).

The formula for $\effect_{1p}^{(1,2)}$ is clear from the path method if $X_1$ is an ancestor of $X_2$ and $X_2$ is an ancestor of $X_p$ in $\Dag$: $\theta_{1p}$ is the effect along all directed paths from 1 to $p$, and we then subtract the effect $\effect_{12}\effect_{2p}$ along the subset of paths that pass through node 2. It is important to note, however, that equation \eqref{eq: RRC double intervention} holds regardless of the causal ordering of $X_1$, $X_2$ and $X_p$.
\begin{remark}\label{remark: joint effect and single intervention effect}
  If $\rvX_1 \in \NdVec_2$, then $\theta_{21}=0$ (see Remark \ref{remark: thetaij=0}) and hence $\theta_{2p}^{(1,2)}=\theta_{2p}$. Similarly, if $\rvX_2 \in \NdVec_1$, then $\theta_{1p}^{(1,2)} = \theta_{1p}$. At least one of these two scenarios must hold due to acyclicity.
\end{remark}
We now generalize Theorem \ref{theorem: soundness of RRC} to $k\ge 2$, yielding a recursive tool to compute total joint effects of any cardinality from lower order effects, and in particular from single intervention effects.

\begin{theorem}\label{theorem: RRC for multiple interventions}
   (Oracle version of RRC for $k\ge 2$) Let $\allvar$ be generated from a linear SEM and let $k\in \{1,\dots,p-1\}$. Then the total effect of $\rvX_i$ $(1\le i\le k)$ on $\rvY$ in a joint intervention on $(\rvX_1,\dots,\rvX_k)$ satisfies:\vspace{-0.05in}
   \begin{align*}
     %\effect_{ip}^{(1,\ldots,k)} =  \effect_{ip}^{(1,\ldots,j-1,j+1,\ldots,k)} - \effect_{ij}^{(1,\ldots,j-1,j+1,\ldots,k)}\effect_{jp}^{(1,\ldots,i-1,i+1,\ldots,k)}.
      \effect_{ip}^{[k]} =  \effect_{ip}^{[k]\setminus \{ j \}} - \effect_{ij}^{[k]\setminus \{ j \}}\effect_{jp}^{[k]\setminus \{ i \}} \qquad \text{for any}\,\,j\in \{1,\dots,k\}\setminus \{i\},\vspace{-0.05in}
   \end{align*} where we use the notation $[k]$ and $[k] \setminus \{j\}$ to denote $(1,\ldots,k)$ and  $(1,\ldots,j-1,j+1,\ldots,k)$, respectively.
\end{theorem}

%%%%%%%%%%%%%%%%%%%%%%%%%%%%%%%%%%%%%%%%%%%%%%%%%%%%%%%%%%%%%%%%%%%%%%%%%%%%%%%%%%%%%%%%%%%%%%%%%%%%%%%%%%%%%%%%%%%%%%%%%%%%%%%

\subsection{Modified Cholesky decompositions (MCD)}\label{subsection: MCD}

Our second OPIN method is based on modifying Cholesky decompositions (MCD) of covariance matrices.
%Throughout this subsection, let $\allvar$ be generated from a \emph{Gaussian} linear SEM characterized by $(\Dag,\errorvec)$, where $\Dag=(\VSet,\ESet,\WMat)$. Let $k\in\{1,\dots,p\}$ denote the number of joint interventions.
The pseudocode is given in Algorithms \ref{algorithm: MCD basic} and \ref{algorithm: MCD}, and the intuition is as follows.
The covariance matrix $\covmat$ of $\allvar$ is given by \vspace{-0.05in}
$$\covmat = (\mathrm{I} -  \WMat^T)^{-1}\Cov(\errorvec) (\mathrm{I} -  \WMat^T)^{-T}.\vspace{-0.05in}$$
Let $\Dag_k=(\VSet, \ESet_k, \WMat_k)$, where $\ESet_k$ is obtained from $\ESet$ by deleting all edges into nodes $\{1,\dots,k\}$ and $\WMat_k$ is obtained from $\WMat$ by setting the columns corresponding to $\rvX_1,\dots,\rvX_k$ equal to zero. $\Dag_k$ is related to the joint intervention on $(\rvX_1,\ldots,\rvX_k)$ as follows. Let $\allvar' = (\rvX_1',\ldots,\rvX_p') ^T$ be generated from the linear SEM $(\Dag_k,\errorvec)$. Then the post intervention joint density of $\allvar$ given the intervention values $(\rvx_1,\ldots,\rvx_k)$ is identical to the conditional distribution of $\allvar'$ given  $(\rvX_1',\ldots,\rvX_k') =  (\rvx_1,\ldots,\rvx_k)$. Let $\Sigma_k$ be the covariance matrix of $\allvar'$, i.e., \vspace{-0.1in}
\begin{align}\label{eq: Sigmak}
  \covmat_k = (\mathrm{I}-\WMat_k^T)^{-1}\Cov(\errorvec)(\mathrm{I}-\WMat_k^T)^{-T}. \vspace{-0.05in}
\end{align}
Then for each $i=1,\ldots,k$, $(\Sigma_k)_{ip} / (\Sigma_k)_{ii}$ equals $\effect_{ip}^{(1,\dots,k)}$, the total effect of $X_i$ on $X_p$ in a joint intervention on $(X_1,\ldots,X_k)$. Hence, we focus on obtaining $\Sigma_k$ from $\Sigma$.

If we knew the causal ordering of the variables, $B$ could be obtained by regressing each variable on its predecessors in the causal ordering, or equivalently, by the generalized Cholesky decomposition. Since $\covmat$ is a positive definite matrix, there exists a unique generalized Cholesky decomposition $(\chol,\diagmat)$, where $\chol \covmat \chol^T = \diagmat$, $\chol$ is a lower triangular matrix with ones on the diagonal, and $\diagmat$ is a diagonal matrix. The first $j-1$ entries of the $j$th row of $L$ correspond to the negative of the regression coefficients in the regression of $\rvX_j$ on $\rvX_1, \dots, \rvX_{j-1}$ \cite{Pourahmadi99}. Hence, if the variables in $\covmat$ are arranged in a causal ordering, the weight matrix $B$ can be obtained from the Cholesky decomposition. Setting the columns of $B$ corresponding to $\rvX_1,\dots,\rvX_k$ equal to zero is therefore equivalent to setting the off-diagonal elements of the rows of $L$ corresponding to $\rvX_1,\dots,\rvX_k$ equal to zero (cf. \cite{BalkePearl94, DrtonEtAl12}). Denoting the resulting matrix by $L_k$, we then obtain $\covmat_k = L_k^{-1} D L_k^{-T}$.

In our set-up, however, we do not know the causal ordering. Instead, we work under the OPIN assumption, knowing only the parent sets of the intervention nodes $\rvX_1,\dots,\rvX_k$. But we can still obtain $\covmat_k$ by an iterative procedure. That is, we first consider a single intervention on $X_1$ to obtain $\covmat_1$. Next, we add the intervention on $X_2$ to obtain $\covmat_2$. After $k$ steps, this yields $\covmat_k$.

The key idea is the following. 
Suppose we wish to construct $\covmat_1$ from $\covmat$, and let $q_1 = | \PaVec_1|$ denote the number of parents of $\rvX_1$. Now order the variables in $\covmat$ such that the first $q_1$ variables correspond to $\PaVec_1$ (in an arbitrary order), the $(q_1+1)$th variable corresponds to $\rvX_1$, and the remaining variables follow in an arbitrary order. Let $(\chol,\diagmat)$ be the generalized Cholesky decomposition of $\Sigma$ with this ordering. Then the first $q_1$ entries of the $(q_1+1)$th row of $L$ contain the negative weights of all edges that are into node $1$ (i.e., these weights are equal to the ones in the true causal weighted DAG). We can then obtain $L_1$ from $L$ by setting the first $q_1$ elements in the $(q_1+1)$th row equal to zero, and use the reverse Cholesky decomposition to construct $\covmat_1 = L_1^{-1} D L_1^{-T}$. 

Repeating this procedure for the other intervention nodes leads to the iterative procedure given in Algorithm \ref{algorithm: MCD basic}, where the matrices in the $j$th step of this algorithm are denoted by $\covmat^{[j]}$, $\chol^{[j]}$ and $\diagmat^{[j]}$. Theorem \ref{theorem: soundness of MCD basic} shows that the output of this algorithm indeed equals $\covmat_k$.

\begin{algorithm}[H]
\caption{}\label{algorithm: MCD basic}
   \begin{algorithmic}[1]
     \REQUIRE $\covmat=\Cov(\allvar)$, parent sets of intervention nodes $(1,\ldots,k )$
     \ENSURE $\covmat^{[k]}$ (which equals $\covmat_k$ as defined in \eqref{eq: Sigmak}, see Theorem \ref{theorem: soundness of MCD basic})
        \STATE set $\covmat^{[0]} = \covmat$;
        \FOR{$j = 1,\ldots,k$}
          \STATE set $\mathbf{\rvW}_j =  \allvar \setminus (\PaVec_j \cup  \{\rvX_j\} ) $;
         \STATE order the variables in $\covmat^{[j-1]}$ as $(\PaVec_j ,\rvX_j,\mathbf{\rvW}_j)$, where the ordering within $\PaVec_j$ and $\mathbf{\rvW}_j$ is arbitrary;
          \STATE obtain the Cholesky decomposition $\chol^{[j-1]} \covmat^{[j-1]} (\chol^{[j-1]})^{T} = \diagmat^{[j-1]}$;
          \STATE obtain $\chol^{[j]}$ from $\chol^{[j-1]}$ by replacing the $(q_j +1)$-th row by $\mathbf{e}_{q_j +1}^T$, where $q_j = |\PaVec_j|$ and $\mathbf{e}_{q_j+1}$ is the $(q_j +1)$-th column of the $p\times p$ identity matrix;
          \STATE set $\covmat^{[j]} = (\chol^{[j]})^{-1} \diagmat^{[j-1]} (\chol^{[j]})^{-T}$;
        \ENDFOR
        \STATE order the variables in $\covmat^{[k]}$ as they were in $\covmat$;
         \RETURN $\covmat^{[k]}$.
   \end{algorithmic}
\end{algorithm}

\begin{theorem}\label{theorem: soundness of MCD basic}
  (Soundness of Algorithm \ref{algorithm: MCD basic}) Let $\allvar$ be generated from a linear SEM and let $k\in \{1,\dots,p-1\}$. Then the output $\covmat^{[k]}$ of Algorithm \ref{algorithm: MCD basic} equals $\covmat_k$ as defined in \eqref{eq: Sigmak}.
\end{theorem}

Since our main goal is to obtain the total joint effect of $(\rvX_1,\ldots,\rvX_k)$ on $\rvX_p$, we do not need to obtain the full covariance matrix $\covmat_k$. Let
\begin{align}\label{eq: impvar}
   \impvar = \{\rvX_{1},\ldots,\rvX_{k}\} \cup \{ \cup_{i=1}^{k} \PaVec_i \} \cup \{ \rvY\}.
\end{align}
Then it suffices to obtain $(\covmat_k)_{\impvar}$, i.e., the sub-matrix of $\covmat_k$ that corresponds to $\impvar$. The proof of Theorem \ref{theorem: soundness of MCD} shows that $(\covmat_k)_{\impvar}$ can be obtained by simply running Algorithm \ref{algorithm: MCD basic} with input matrix $\Cov(\impvar)$. This simplification is important in sparse high-dimensional settings, where the full covariance matrix $\Sigma$ can be very large and difficult to estimate, while the sub-matrix $\Cov(\impvar)$ is small. We can then compute the total joint effect of $(\rvX_1,\dots,\rvX_k)$ on $\rvY$ as indicated in Algorithm \ref{algorithm: MCD}.

\begin{algorithm}[H]
  \caption{MCD oracle}
  \label{algorithm: MCD}
  \begin{algorithmic}[1]
  \REQUIRE $\covmat = \Cov (\impvar)$ (see \eqref{eq: impvar}), parent sets of intervention nodes $(1,\ldots,k )$
  \ENSURE $(\effect_{1p}^{\prime (1,\ldots, k)},\ldots,\effect_{kp}^{\prime (1,\ldots, k)})^T$ (which equals $\effectvec^{(1,\dots,k)}$ as defined in Definition \ref{definition: total causal effect}, see Theorem \ref{theorem: soundness of MCD})
      \STATE run Algorithm \ref{algorithm: MCD basic} with input $\covmat$ to obtain $\covmat^{[k]}$;
      \STATE for $i=1,\ldots,k$, obtain $\effect_{ip}^{\prime (1,\ldots, k)} = 1_{\{\rvX_p \notin \PaVec_{i} \}}\covmat^{[k]}_{\rvx_i\rvx_p} / \covmat^{[k]}_{\rvx_i\rvx_i}$, where $\covmat^{[k]}_{\rvx_i\rvx_j}$ is the entry of $\covmat^{[k]}$ that corresponds to $(\rvX_i,\rvX_j)$; \label{line: total joint effect} \label{line: unadjusted regression}
      \RETURN $(\effect_{1p}^{\prime (1,\ldots, k)},\ldots,\effect_{kp}^{\prime (1,\ldots, k)})^T$.
  \end{algorithmic}
\end{algorithm}

\begin{theorem}\label{theorem: soundness of MCD} (Soundness of MCD oracle)
 Let $\allvar$ be generated from a linear SEM and let $k\in \{1,\dots,p-1\}$. Then the output of Algorithm \ref{algorithm: MCD} equals $\effectvec^{(1,\dots,k)}$, the total joint effect of $(\rvX_1,\dots,\rvX_k)$ on $\rvY$.
\end{theorem}

%The Gaussian assumption in Theorems \ref{theorem: soundness of MCD basic} and \ref{theorem: soundness of MCD} is not necessary. This assumption is only used in the proof of Lemma \ref{A2-lemma: for soundness of MCD basic} of \cite{NandyMaathuisRichardson14b}, since this simplified our proof considerably.

The term $1_{\{\rvX_p \notin \PaVec_{i} \}}$ in line \ref{line: total joint effect} of Algorithm \ref{algorithm: MCD} is not necessary for the oracle version, since $\covmat_{x_ix_p}^{[k]} = 0$ if $X_p \in \PaVec_i$. However, it does make a difference in the sample version, when we use the sample covariance matrix instead of the true covariance matrix. %, but we have an alternative proof for Lemma \ref{lemma: for soundness of MCD basic} that allows arbitrary mean zero distributions for the $\epsilon_i$'s.
Section \ref{A2-section: an illustration of the MCD algorithm} of \cite{NandyMaathuisRichardson14b} contains a detailed example where MCD is applied to the weighted DAG in Figure \ref{fig: path method}.

\begin{remark}\label{remark: hidden variable}
%Theorem \ref{theorem: soundness of MCD} shows that the MCD algorithm is sound if parent sets of the intervention nodes are correctly specified. 
Soundness of RRC and MCD may still hold even when the parent sets are not correctly specified. We show this in Section \ref{A2-section: hidden variable examples} of \cite{NandyMaathuisRichardson14b} with two examples with incorrectly specified parent sets. In Example \ref{A2-example: hidden variable example 1} of \cite{NandyMaathuisRichardson14b}, both RRC and MCD produce the correct output, while only MCD produces the correct output in Example \ref{A2-example: hidden variable example 2} of \cite{NandyMaathuisRichardson14b}. The latter shows that RRC and MCD are indeed two different approaches, although their outputs are identical in the oracle versions when the parent sets are correctly specified.
\end{remark}

%%\begin{remark}
%\textcolor{red}{Note that Theorems \ref{theorem: RRC for multiple interventions} and \ref{theorem: soundness of MCD} imply that the outputs of the oracle versions of RRC and MCD are identical. However, RRC and MCD represent two different functions of the covariance matrix and hence lead to two different plug-in estimators (when $k\geq 2$) in the sample versions defined below.}
%%\end{remark}

%%%%%%%%%%%%%%%%%%%%%%%%%%%%%%%%%%%%%%%%%%%%%%%%%%%%%%%%%%%%%%%%%%%%%%%%%%%%%%%%%%%%%%%%%%%%%%%%%%%%%%%%%%%%%%%%%%%%%%%%%%%%%%%

\subsection{Sample versions}\label{subsection: sample versions}

Suppose that we have $n$ i.i.d.\ observations of $\allvar$, where $\allvar$ is generated from a linear SEM characterized by $(\Dag,\errorvec)$.

We first define an adjusted regression estimator for $\effect_{1p}$, the total causal effect of $\rvX_1$ on $\rvY$  \citep{MaathuisKalischBuehlmann09} (cf. equation  \eqref{eq: causal effect}):
\begin{align}\label{eq: causal effect estimate}
  \hat \effect_{1p} =   \left\{ \begin{array}{ll} 0 & \text{if $\rvY \in \PaVec_1$} \\  \hat \beta_{1p | \PaVec_1} & \text{otherwise} \end{array} \right.,
\end{align}
where $\hat \beta_{1p | \PaVec_1}$ is the sample regression coefficient of $\rvX_1$ in the linear regression $\rvY \sim \rvX_1 + \PaVec_1$.
%\begin{remark}\label{remark: hard zeros}
%Note that if $\rvX_p$ is a parent of $\rvX_1$, then both $\effect_{1p}$ and its estimate $\hat \effect_{1p}$ are zero. If $\rvX_p$ is a non-descendant but not a parent of $\rvX_1$, then $\hat \effect_{1p}$ is a sample dependent estimate of the true value $\effect_{1p} = 0$.
%\end{remark}

Next, we define sample versions of RRC and MCD.
\begin{definition}\label{definition: RRC estimator} (RRC estimator)
   Let $k\in \{1,\dots,p-1\}$. The RRC estimator for the total effect of $\rvX_i$ $(1\le i\le k)$ on $\rvY$ in a joint intervention on $(\rvX_1,\dots,\rvX_k)$ is defined recursively as follows
   (cf. Theorem \ref{theorem: RRC for multiple interventions}):
   \begin{align*}
      \hat{\effect}_{ip}^{[k]} =   \left\{
        \begin{array}{ll}
            0 & \text{if $\rvY \in \PaVec_i$} \\
            \hat{\effect}_{ip}^{[k]\setminus \{j\}} - \hat{\effect}_{ik}^{[k] \setminus \{j\}}\hat{\effect}_{kp}^{[k]\setminus \{i\}} & \text{otherwise}
        \end{array}
        \right.,
   \end{align*}
   where the adjusted regression estimator $\hat{\effect}^{(i)}_{ij} = \hat{\effect}_{ij}$ is defined in \eqref{eq: causal effect estimate} and we fix $j = \max([k] \setminus \{i\})$.
\end{definition}

Thus, the effects of multiple interventions can be estimated from single intervention effects, where the latter can be estimated from single intervention experiments or from observational data and an IDA-like method.

\begin{definition}\label{definition: MCD estimator} (MCD estimator)
   Let $k\in \{1,\dots,p-1\}$.  Let $\hat{\covmat}$ be the sample covariance matrix of $\impvar$ (see \eqref{eq: impvar}). Then the MCD estimator $\tilde{\boldsymbol\theta}_p^{(1,\dots,k)}= (\tilde \theta_{1p}^{(1,\dots,k)}, \dots,\tilde \theta_{kp}^{(1,\dots,k)})^T$ for the total joint effect of $(\rvX_1,\dots,\rvX_k)$ on $\rvY$ is the output of Algorithm \ref{algorithm: MCD} when $\hat{\covmat}$ and parent sets of $(1,\dots,k)$ are used as input.
\end{definition}

The MCD estimator for $k=1$ simply equals adjusted regression:
\begin{theorem}\label{theorem: RRC and MCD for single interventions} (MCD estimator for single interventions)
   Let $\hat{\effect}_{1p}$ be as in \eqref{eq: causal effect estimate} and $\tilde{\effect}_{1p}$ as in Definition \ref{definition: MCD estimator}. Then $\tilde{\effect}_{1p} = \hat{\effect}_{1p}$.
\end{theorem}

Finally, we note that the RRC estimator for $k \geq 3$ and the MCD estimator for $k \geq 2$ generally depend on the ordering of $\rvX_1,\ldots,\rvX_k$. However, using different orderings in simulations showed very little difference for $k = 2$ or $3$, especially when the underlying causal DAG was sparse.

%% file: texfiles/theoretical_properties.tex
\section{Asymptotic distributions of RRC and MCD}\label{section: theoretical properties}

We now derive the asymptotic distributions of the RRC and MCD estimators under the OPIN assumption.
For simplicity, we limit ourselves to the case $k=2$.

Assume that we have $n$ i.i.d.\ observations of $\allvar$, where $\allvar$ is generated from a linear SEM characterized by $(\Dag,\errorvec)$. Moreover, in this section we assume that $\Exp[\error_i^4] < \infty$ for all $i=1,\ldots,p$. Let $\impvar := \{\rvX_{1},\rvX_{2}\} \cup \{ \cup_{i=1}^{2} \PaVec_i \} \cup \{ \rvY\}$ and $|\impvar| = q$. Let $\covmat_{q\times q} := \Cov(\impvar)$ and let $\sampcovmat$ denote the corresponding sample covariance matrix. The half-vectorization, $\halfvec(A)$, of a symmetric $q \times q$ matrix $A$ is the column vector in $\mathbb{R}^{q(q+1)/2}$ obtained by vectorizing the lower triangular part of $A$. The derivative of a vector-valued differentiable function $\mathbf{y}(\mathbf{x}) = (y_1(\mathbf{x}),\ldots,y_r(\mathbf{x}))^T$ with respect to $\mathbf{x} = (x_1,\ldots,x_s)^T$ is denoted by the $r \times s$ matrix $\frac{\partial \mathbf{y}}{\partial \mathbf{x}}$ whose $(i,j)$-th entry is equal to $\frac{\partial y_i}{\partial x_j}$.

Since all fourth moments of the variables in $\impvar$ are finite, the multivariate central limit theorem implies
\vspace{-0.1in}
\begin{equation}\label{eq: asymptotic distribution of covmat}
\sqrt{n}(\halfvec (\sampcovmat) - \halfvec(\covmat) ) \stackrel{d}{\longrightarrow} \mathcal{N}(\mathbf{0}, \Gamma), \vspace{-0.05in}
\end{equation}
where $\Gamma := \Cov(\halfvec (\impvar \impvar^T))$. We will use (\ref{eq: asymptotic distribution of covmat}) and the multivariate delta-method to derive the asymptotic distributions of RRC and MCD.

%%%%%%%%%%%%%%%%%%%%%%%%%%%%%%%%%%%%%%%%%%%%%%%%%%%%%%%%%%%%%
%\subsection{Asymptotic distributions}\label{subsection: asymptotic distributions}

%We denote the RRC estimator based on $n$ i.i.d.\ observations by $\hat{\boldsymbol\theta}^{(1,2)}_p = (\hat{\effect}_{1p} - \hat{\effect}_{12}\hat{\effect}_{2p},\hat{\effect}_{2p} - \hat{\effect}_{21}\hat{\effect}_{1p})^T$, suppressing the dependence on $n$.

\begin{theorem}\label{theorem: asymptotic distribution of RRC} (Asymptotic distribution of RRC) Assume that $\allvar$ is generated from a linear SEM characterized by $(\Dag,\errorvec)$, where $\Exp[\error_i^4] < \infty$ for all $i=1,\ldots,p$. Moreover, assume that $\{\rvX_{1}, \rvX_2,\rvX_p\} \cap (\PaVec_{1} \cup \PaVec_2) = \emptyset$. Then\vspace{-0.05in}
 $$ \sqrt{n}(\hat{\boldsymbol\effect}_{p}^{(1,2)} - \effectvec^{(1,2)} ) \stackrel{d}{\longrightarrow} \mathcal{N}\left(\mathbf{0}, F \Lambda \Gamma \Lambda^TF^T \right),\vspace{-0.05in}$$ where $\Gamma = \Cov(\halfvec (\impvar \impvar^T))$, $\Lambda = \frac{\partial \boldsymbol\theta}{\partial \halfvec (\covmat)} $ with ${\boldsymbol\theta} := ({\effect}_{1p},{\effect}_{2p},{\effect}_{12},{\effect}_{21})^T$, and $F = \frac{\partial \effectvec^{(1,2)}}{\partial \boldsymbol\theta}$. An explicit expression of the ${4\times q(q+1)/2}$ matrix $\Lambda$ is given in Proposition \ref{A2-proposition: asymptotic distribution of RRC} of \cite{NandyMaathuisRichardson14b}.
 \end{theorem}

By definition, $\hat{\effect}_{ij}=0$ if $\rvX_{j} \in \PaVec_{i}$ for any $i = 1,2$, $j=1,2,p$ and $j \neq i$. Hence, the cases excluded from Theorem \ref{theorem: asymptotic distribution of RRC} can be handled trivially.

To obtain the asymptotic distribution of MCD, we first derive the asymptotic distribution of $\halfvec (\sampcovmat^{[2]})$, where $\sampcovmat^{[2]}$ is the output of Algorithm \ref{algorithm: MCD basic} applied to $\sampcovmat$. Without loss of generality, we assume that the variables in $\covmat$ are ordered as $\PaVec_1,\rvX_1,\impvar \setminus (\PaVec_1 \cup X_1)$, where the variable in $\PaVec_1$ and $\impvar \setminus (\PaVec_1 \cup \rvX_1)$ are ordered arbitrarily. Moreover, for simplicity of notation we omit line 9 of Algorithm \ref{algorithm: MCD basic}, so that the variables in $\sampcovmat^{[2]}$ are ordered as $(\PaVec_2, \rvX_2, \impvar \setminus \PaVec_2 \cup \rvX_2)$.
 
%For simplicity, we assume that the variables in $\sampcovmat^{[2]}$ and $\covmat^{[2]}$ are ordered as $(\PaVec_2,\rvX_2,\impvar \setminus (\PaVec_2 \cup \rvX_2))$, where variables in $\PaVec_2$ and $\impvar \setminus (\PaVec_2 \cup \rvX_2)$ are ordered arbitrarily. Let $\covmat^{[1]}$ and $\sampcovmat^{[1]}$ denote the one-step modification of $\covmat$ and $\sampcovmat$ through Algorithm \ref{algorithm: MCD basic}. Thus variables in $\covmat^{[1]}$, $\sampcovmat^{[1]}$ are ordered as $(\PaVec_1,\rvX_1,\impvar \setminus (\PaVec_1 \cup \rvX_1))$ and without loss of generality we assume that the variables in $\covmat$, $\sampcovmat$ have the same ordering.

Let $P$ be the $q \times q$ permutation matrix such that variables in $P\sampcovmat^{[1]} P^T$ are ordered as in $\sampcovmat^{[2]}$. %$(\PaVec_2,\rvX_2,\impvar \setminus (\PaVec_2 \cup \rvX_2))$. 
Let $\Pi$ be the matrix such that for any $q \times q$ symmetric matrix $A$, $\halfvec(P A P^T) = \Pi ~\halfvec (A)$. We define $\Lambda^{[1]} =  \frac{\partial \halfvec(\covmat^{[1]})}{\partial \halfvec (\covmat)}$ and $ \Lambda^{[2]} =   \frac{\partial \halfvec(\covmat^{[2]})}{\partial \halfvec (P\covmat^{[1]} P^T)}$.

\begin{theorem}\label{theorem: asymptotic distribution of the modified covmat} (Asymptotic distribution of $\sampcovmat^{[2]}$) Assume that $\allvar$ is generated from a linear SEM characterized by $(\Dag,\errorvec)$, where $\Exp[\error_i^4] < \infty$ for all $i=1,\ldots,p$. Then \vspace{-0.05in}
$$\sqrt{n} (\halfvec(\sampcovmat^{[2]}) - \halfvec (\covmat^{[2]}) ) \stackrel{d}{\longrightarrow} \mathcal{N}\left(\mathbf{0},  \Gamma^{[2]} \right), \vspace{-0.05in}$$
where $\Gamma^{[2]} := \Lambda^{[2]}\Pi\Lambda^{[1]}\Gamma \Lambda^{[1]T}\Pi^T \Lambda^{[2]T}$ with $\Gamma = \Cov(\halfvec (\impvar \impvar^T))$. An explicit expression for $\Lambda^{[1]}$ is given in Section \ref{A2-section: Supplement to Section 4} of \cite{NandyMaathuisRichardson14b}, and an expression for $\Lambda^{[2]}$ can be obtained analogously.
\end{theorem}

%We denote the MCD estimator based on $n$ i.i.d.\ observations by $\tilde{\boldsymbol\effect}_{p}^{(1,2)}$, suppressing its dependence on $n$. 
The following corollary follows directly from Theorem \ref{theorem: asymptotic distribution of the modified covmat} and the multivariate delta-method.
\begin{corollary}\label{corollary: asymptotic distribution of MCD} (Asymptotic distribution of MCD) Assume that $\allvar$ is generated from a linear SEM characterized by $(\Dag,\errorvec)$, where $\Exp[\error_i^4] < \infty$ for all $i=1,\ldots,p$. Moreover, assume that $\rvX_p \notin \PaVec_{1} \cup \PaVec_2$. Then \vspace{-0.05in} $$ \sqrt{n}(\tilde{\boldsymbol\effect}_{p}^{(1,2)} - \effectvec^{(1,2)} ) \stackrel{d}{\longrightarrow} \mathcal{N}\left(\mathbf{0}, H\Gamma ^{[2]} H^T \right),\vspace{-0.05in}$$ where $\Gamma^{[2]}$ is defined in Theorem \ref{theorem: asymptotic distribution of the modified covmat}, $\effectvec^{(1,2)} = \left(\covmat^{[2]}_{\rvx_1\rvx_p}/\covmat^{[2]}_{\rvx_1\rvx_1},~ \covmat^{[2]}_{\rvx_2\rvx_p}/\covmat^{[2]}_{\rvx_2\rvx_2}\right)^T$ and $H = \frac{\partial \effectvec^{(1,2)}}{\partial \halfvec (\covmat^{[2]})}$.
\end{corollary}

Since the formulas in Theorem \ref{theorem: asymptotic distribution of RRC} and Corollary \ref{corollary: asymptotic distribution of MCD} are not easily comparable, we computed them numerically for various settings in Section \ref{A2-section: asymptotic comparison} of \cite{NandyMaathuisRichardson14b}. We found that no estimator dominates the other in terms of asymptotic variance, but that RRC seems to have a smaller asymptotic variance in most cases.

%%%%%%%%%%%%%%%%%%%%%%%%%%%%%%%%%%%%%%%%%%%%%%%%%%%%%%%%%%%%%

%% file: texfiles/parents_sets_from_cpdag.tex
\section{Extracting possible parent sets from a CPDAG}\label{section: parent sets from cpdag}

Recall that we introduced the OPIN assumption in Section \ref{section: multiple interventions using parent sets} as a stepping stone for the scenario where we have no information on the underlying causal DAG. We will now consider this more general scenario: $\allvar$ is generated from a linear SEM, and we only know the observational distribution of $\allvar$.

%\textcolor{red}{As mentioned before, the causal DAG is not identifiable, except for some special cases \cite{ShimizuEtAl06-JMLR, PetersBuhlmann14}. Instead,}
As in IDA we can first estimate the CPDAG $\mathcal C$ of the unknown underlying causal DAG. Conceptually, we could then list all DAGs in the Markov equivalence class described by $\mathcal C$ (see, e.g., \cite{DorTarsi92}). Suppose that the Markov equivalence class consists of $m$ DAGs $\{\Dag_1,\dots,\Dag_m\}$. For each $\Dag_j$, $j=1,\dots,m$, we could determine the parent sets of the intervention nodes $(1,\dots,k)$, denoted by the ordered set $\PaVec(\Dag_j) = (\PaVec_1(\Dag_j),\dots,\PaVec_k(\Dag_j))$. All possible jointly valid parent sets of $(1,\dots,k)$ are then
\vspace{-0.05in}
\begin{align}\label{eq: PAall}
   \mathcal{PA}_{all} = \{\PaVec(\Dag_j) \st j =1,\ldots,m\}. \vspace{-0.05in}
\end{align}
We could then apply the RRC and MCD algorithms, using each of the possible jointly valid parent sets $\PaVec(\Dag_j)$, $j=1,\dots,m$, to obtain the multiset of possible total joint effects \vspace{-0.05in}
\begin{align}\label{eq: Theta*}
    \Theta_{p}^{*(1,\dots,k)} & = \{ \effectvec^{(1,\dots,k)}(\PaVec(\Dag_j)) \st j=1,\dots,m \}. \vspace{-0.05in}
\end{align}

However, listing all DAGs is computationally expensive and does not scale well to large graphs.  In this section, our aim is to develop efficient ways to find the jointly valid parent sets of $(1,\dots,k)$, where parent sets are called \emph{jointly valid} if there exists a DAG in the Markov equivalence class of $\mathcal{C}$ with this particular combination of parent sets.

In \cite{MaathuisKalischBuehlmann09}, the authors defined a so-called ``locally valid" parent set of a node and showed that any locally valid parent set of a single intervention node is also a valid parent set. All locally valid parent sets of node $i$ can be obtained efficiently by %taking the parents of $i$ in $\mathcal{C}$ and then 
orienting only those undirected edges in $\mathcal{C}$ which contain node $i$ as an endpoint, without creating new v-structures with $i$ as a collider, and then taking all resulting parent sets of $i$. As an easy extension of the method of \cite{MaathuisKalischBuehlmann09} to multiple interventions, one could try to obtain jointly valid parent sets by taking all combinations of the locally valid parent sets of all intervention nodes. However, in Example \ref{ex: local method problem} we show that this approach may generate parent sets that are not jointly valid. In other words, local validity of each parent set is necessary, but not sufficient for joint validity.

\begin{example}\label{ex: local method problem}
   \begin{figure}[!ht]
   \vspace{-0.1in}
     \centering
     \begin{subfigure}[b]{0.24\textwidth}
       \centering
     \includegraphics[width=0.8\textwidth]{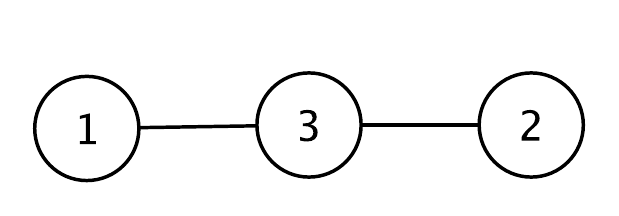}
     \caption{}
     \label{subfig: a simple CPDAG}
     \end{subfigure}
       \begin{subfigure}[b]{0.72\textwidth}
       \centering
     \includegraphics[width=0.8\textwidth]{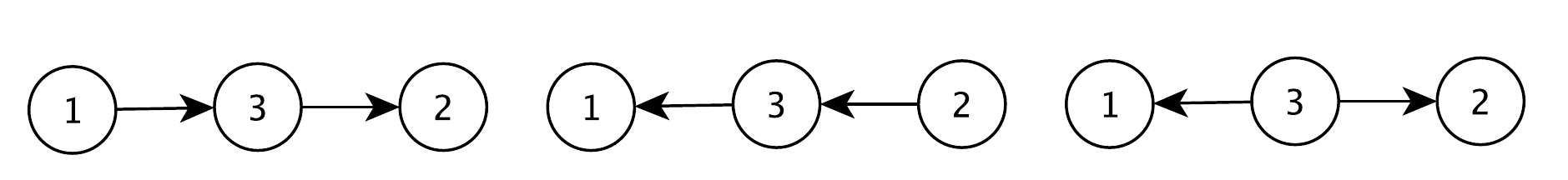}
     \caption{}
     \label{subfig: all DAGs}
     \end{subfigure}
     \caption{(a) A CPDAG $\mathcal{C}$ and (b) all DAGs in its Markov equivalence class. }
     \label{fig: a simple CPDAG}
     \vspace{-0.1in}
   \end{figure}
   Consider the CPDAG $\mathcal{C}$ in Figure \ref{subfig: a simple CPDAG}. There are three DAGs that belong to the Markov equivalence class represented by $\mathcal{C}$ (Figure \ref{subfig: all DAGs}). Thus, all jointly valid sets of parent sets of $(1,2)$ are $(\emptyset,\{3\})$, $ (\{3\}, \emptyset)$ and $ (\{3\}, \{3\})$. However, both $\emptyset$ and $\{3\}$ are locally valid parent sets of vertices $1$ and $2$ in $\mathcal{C}$. Hence, all combinations of locally valid parent sets of nodes $1$ and $2$ include the additional ordered set $(\emptyset, \emptyset)$. The latter is not jointly valid, since it corresponds to the DAG $1\rightarrow 3 \leftarrow 2$, which is not in the Markov equivalence class represented by $\mathcal{C}$ due to the additional v-structure.
\end{example}

%\subsection{Semi-local algorithm for extracting parent sets}

We now propose a semi-local algorithm for extracting all jointly valid parent sets, using the following graph-theoretic property of a CPDAG: no orientation of edges not oriented in a CPDAG $\mathcal{C}$ can create a directed cycle or a new v-structure which includes at least one edge that was oriented in $\mathcal{C}$ (see the proof of Theorem 4 in \cite{Meek95}).

 Let $\mathcal{C}_{undir}$ and $\mathcal{C}_{dir}$ be the subgraphs on all vertices of $\mathcal{C}$ that consist of all undirected and all directed edges of $\mathcal{C}$, respectively.  Then $(\PaVec_1^\prime,\ldots,\PaVec_k^\prime)$ is a jointly valid parent set of the intervention nodes with respect to $\mathcal{C}_{undir}$ if and only if $(\PaVec_1^\prime\cup\PaVec_1(\mathcal{C}_{dir}),\ldots,\PaVec_k^\prime\cup\PaVec_k(\mathcal{C}_{dir}))$ is a jointly valid parent set of the intervention nodes with respect to $\mathcal{C}$. Typically, $\mathcal{C}_{undir}$ consists of several connected components, and these can be considered independently of each other. Since we only have to consider components that contain an intervention node, we have to work with at most $k$ components (which are typically much smaller than $\mathcal{C}$), and this gives a large computational advantage. The algorithm is given in pseudocode in Algorithm \ref{algorithm: semi-local algorithm} and illustrated Example \ref{ex: semi-local method}.
% It can be expected that $\mathcal{C}_{undir}$ has more disconnected components than $\mathcal{C}$, and this gives a large computational advantage, since one can work with each component \textcolor{red}{(that contains at least one intervention node)} independently.

\begin{algorithm}[H]
   \caption{Extracting jointly valid parents sets of intervention nodes from a CPDAG}\label{algorithm: semi-local algorithm}
   \begin{algorithmic}[1]
   \REQUIRE CPDAG $\mathcal{C}$, intervention nodes $(1,\ldots, k )$
   \ENSURE A multiset $\mathcal{PA}_{s\ell}$ (which is equivalent to $\mathcal{PA}_{all}$ by Theorem \ref{theorem: soundness of the semi-local algorithm})
   \STATE obtain $\mathcal{C}_{undir}$ and $\mathcal{C}_{dir}$ from $\mathcal{C}$;
   \STATE let $ \mathcal{C}_{1}, \ldots, \mathcal{C}_{s}$ be the connected components of $\mathcal{C}_{undir}$ that contain at least one intervention node (note $s\leq k$); \label{line: connected components}
   \STATE for $i=1,\dots,s$, let $\mathcal{PA}_{i} $ be the multiset of all jointly valid parent sets of the intervention nodes in $\mathcal{C}_i$, obtained by constructing all DAGs in the Markov equivalence class described by $\mathcal{C}_i$; \label{line: parent sets in a chordal component}
   \STATE form $\mathcal{PA}_{undir}(1,\ldots,k)$ by taking all possible combinations of $\mathcal{PA}_{1},\ldots,\mathcal{PA}_{s}$ (as in Example \ref{ex: semi-local method});
   \STATE $\mathcal{PA}_{s\ell} \! \leftarrow \!\{(\PaVec_1^\prime\cup\PaVec_1(\mathcal{C}_{dir}),\ldots,\PaVec_k^\prime\cup\PaVec_k(\mathcal{C}_{dir}))\!\! \st \!\!(\PaVec_1^\prime,\ldots,\PaVec_k^\prime)\!\in\!\mathcal{PA}_{undir}(1,\ldots,k) \}$;
   \RETURN $\mathcal{PA}_{s\ell}$.
   \end{algorithmic}
\end{algorithm}

\begin{example}\label{ex: semi-local method}
   \begin{figure}[!ht]
   %\vspace{-0.1in}
     \centering
     \begin{subfigure}[b]{0.3\textwidth}
      \centering
     \includegraphics[width=0.8\textwidth]{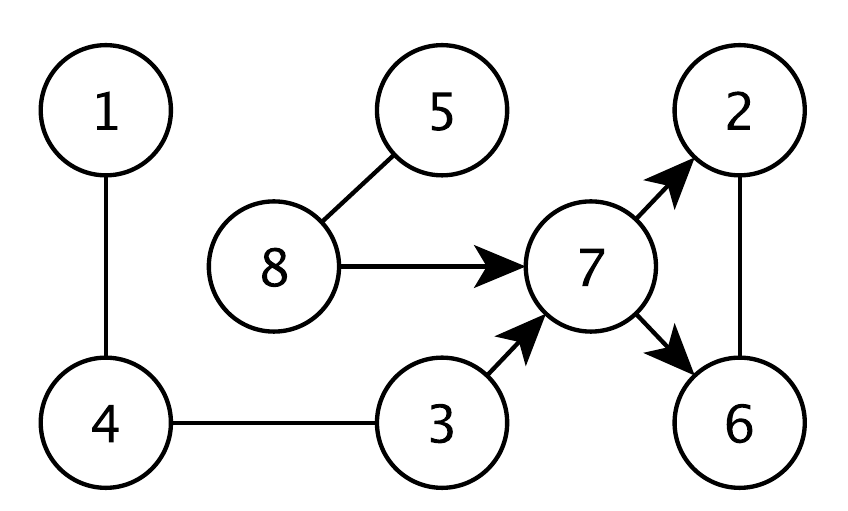}
     \caption{}
     \label{subfig: cpdag}
     \end{subfigure}
       \begin{subfigure}[b]{0.3\textwidth}
       \centering
     \includegraphics[width=0.8\textwidth]{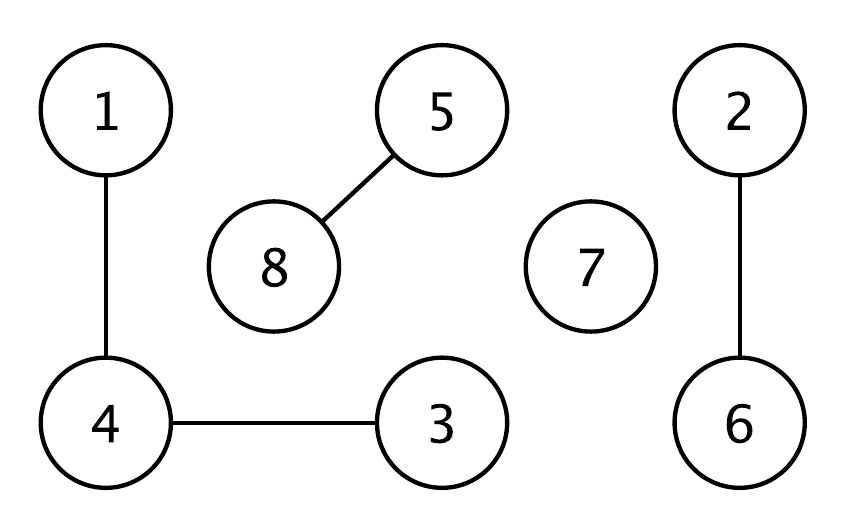}
     \caption{}
     \end{subfigure}
       \begin{subfigure}[b]{0.3\textwidth}
        \centering
     \includegraphics[width=0.8\textwidth]{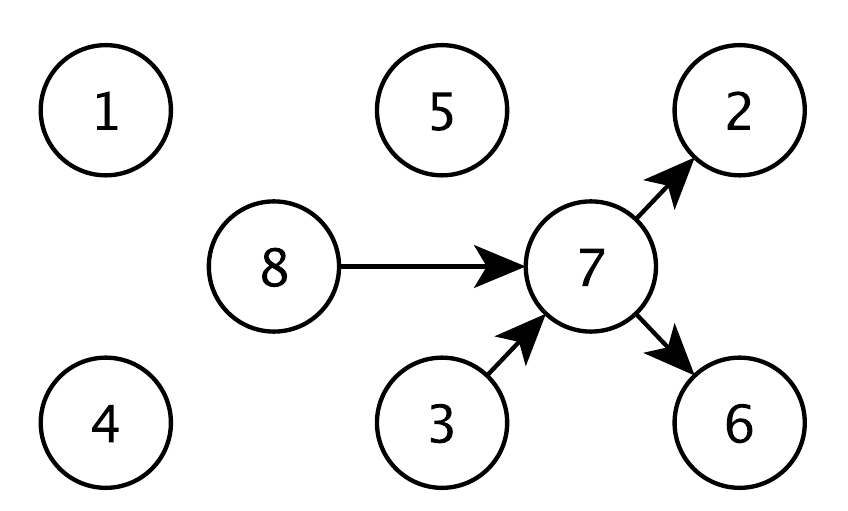}
     \caption{}
     \end{subfigure}
     \caption{(a) A CPDAG $\mathcal{C}$ and its corresponding subgraphs (b) $\mathcal{C}_{undir}$ and (c) $\mathcal{C}_{dir}$. }
     \label{fig: CPDAG example}
     \vspace{-0.1in}
   \end{figure}
Consider the CPDAG $\mathcal{C}$ in Figure \ref{fig: CPDAG example}, together with its corresponding subgraphs $\mathcal{C}_{undir}$ and $\mathcal{C}_{dir}$.  We assume that the intervention nodes are $(1,2,3)$. Note that $\mathcal{C}_{undir}$ contains four connected components and two of them contain at least one intervention node, namely $\mathcal{C}_1: 1 - 4 - 3$ and $\mathcal{C}_2:  2 - 6$. We first consider $\mathcal{C}_1$. The multiset of all possible jointly valid parent sets of $(1,3)$ with respect to $\mathcal{C}_1$ can be obtained by creating all possible DAGs in the Markov equivalence class described by $\mathcal C_1$. This yields $\mathcal{PA}_1 = \{ (\emptyset,\{4\}), ( \{ 4\}, \emptyset ),( \{4\}, \{4\}) \}$ (see Example \ref{ex: local method problem}). Next, considering $\mathcal{C}_2$ we find that the multiset of possible parent sets of $2$ with respect to $\mathcal{C}_2$ is $\mathcal{PA}_2 =\{ \emptyset, \{6\}\}$. By taking all combinations of $\mathcal{PA}_1$ and $\mathcal{PA}_2$, we obtain
   \begin{align*}
      \mathcal{PA}_{undir}(1,2,3) = \{ & (\emptyset, \emptyset, \{4\}),  (\emptyset, \{6\} ,\{4\}), (\{ 4\},\emptyset ,\emptyset),\\
                                       & (\{ 4\},\{6\},\emptyset), (\{4\},\emptyset,\{4\}), (\{4\},\{6\},\{4\}) \}.
   \end{align*}

   Furthermore, $\PaVec_1(\mathcal{C}_{dir}) = \emptyset$, $\PaVec_{2}(\mathcal{C}_{dir}) = \{7\}$, and $\PaVec_3(\mathcal{C}_{dir}) = \emptyset$. Combining this with $\mathcal{PA}_{undir}(1,2,3)$ yields
   \begin{align*}
      \mathcal{PA}_{s\ell} = \{ & (\emptyset,\{7\}, \{4\}),  (\emptyset, \{6,7\} ,\{4\}), (\{ 4\},\{7\} ,\emptyset),\\
                                    & (\{ 4\},\{6,7\},\emptyset), (\{4\},\{7\},\{4\}),  (\{4\},\{6,7\},\{4\}) \}.
   \end{align*}
   Finally, $\mathcal{PA}_{all}$ (see equation \ref{eq: PAall}) is a multiset of size $12$, where each element of $\mathcal{PA}_{s\ell}$ occurs twice due to the two possible orientations of the edge $5-8$.
\end{example}

We say that two multisets $A$ and $B$ are \emph{equivalent} (up to ratios) if (i) $A \stackrel{\mathrm{set}}{=} B$, i.e., the set of all distinct elements of $A$ is equal to that of $B$, and (ii) the ratios of multiplicities of any two elements in $A$ is equal to their ratio of multiplicities in $B$. For example, $A = \{a,a,b\}$ and $B = \{a,a,a,a,b,b \}$ are equivalent multisets.

In Example \ref{ex: semi-local method}, we saw that $\mathcal{PA}_{all}$ and $\mathcal{PA}_{s\ell}$ were equivalent. Theorem \ref{theorem: soundness of the semi-local algorithm} shows that this equivalence holds in general.

\begin{theorem}\label{theorem: soundness of the semi-local algorithm} (Soundness of Algorithm \ref{algorithm: semi-local algorithm})
   Let $\mathcal{PA}_{s\ell}$ be the output of Algorithm \ref{algorithm: semi-local algorithm} and $\mathcal{PA}_{all}$ be as in \eqref{eq: PAall}. Then $\mathcal{PA}_{s\ell}$ and $\mathcal{PA}_{all}$ are equivalent.
\end{theorem}
We note that the local method used for single interventions in IDA does not yield a multiset that is equivalent to the global method of listing all the DAGs. The distinct elements of the two resulting multisets are the same, but the local method loses the multiplicity information. Thus, if multiplicity information is important, the semi-local algorithm proposed here can also be used in IDA.

In Section \ref{A2-section: comp time comparison} of \cite{NandyMaathuisRichardson14b}, we compare the computation time of the semi-local Algorithm \ref{algorithm: semi-local algorithm} for single interventions to that of the local algorithm of IDA. Numerical results show that the computation times are comparable, except for a few extreme cases. Note that the computation time for obtaining all DAGs in the Markov equivalence class described by a component $\mathcal C_i$ (line 3 of Algorithm \ref{algorithm: semi-local algorithm}) is exponential in the size of the component. This step thus becomes infeasible for large components (larger than, say, $12$ nodes). In our simulations this occurred only in approximately $0.1\%$ of the cases. In these rare cases, we recommend to obtain $\mathcal{PA}_i$ in line \ref{line: parent sets in a chordal component} of Algorithm \ref{algorithm: semi-local algorithm} by combining all locally valid parent sets of the intervention nodes in $\mathcal C_i$ (as in Example \ref{ex: local method problem}). This generally leads to a superset of jointly valid parent sets, and hence to a superset of possible causal effects.

 %Note that line \ref{line: connected components} of Algorithm \ref{algorithm: semi-local algorithm} ensures that, on average, computational complexity of the most challenging part (line \ref{line: parent sets in a chordal component}) of Algorithm \ref{algorithm: semi-local algorithm} can increase at most linearly with $k$.

We discuss a sample version of Algorithm \ref{algorithm: semi-local algorithm} in Section \ref{A2-section: sample version of the semi-local algorithm} of \cite{NandyMaathuisRichardson14b}. This sample version addresses an additional issue, namely that some undirected components of the estimated partially directed graph may not describe a Markov equivalence class of DAGs. In such cases, we again recommend combining all locally valid parent sets.

%\subsection{Sample version}\label{subsection: semi local algorithm sample version}

%% file: texfiles/multiple_interventions_from_data.tex
\section{Estimation from observational data}\label{section: estimation from observational data}

We now combine the methods of Sections \ref{section: multiple interventions using parent sets} and \ref{section: parent sets from cpdag}. Given a true CPDAG $\mathcal C$ and covariance matrix $\Sigma$, we define
the following (oracle) multiset of possible total joint effects of $(\rvX_1,\dots,\rvX_k)$ on $\rvY$: \vspace{-0.05in}
\begin{align}\label{eq: Theta}
    \Theta_p^{(1,\ldots,k)} & := \{ \effectvec^{(1,\ldots,k)}(\PaVec') \st \PaVec' \in \mathcal{PA}_{s\ell} \},
 \end{align}
where $\mathcal{PA}_{s\ell}$ is the output of Algorithm \ref{algorithm: semi-local algorithm} applied to the true CPDAG $\mathcal{C}$ and the intervention nodes $(1,\ldots,k)$, and $\effectvec^{(1,\ldots,k)}(\PaVec^{\prime})$ is the vector of total joint effects of $(\rvX_{1},\ldots,\rvX_{k})$ on $\rvY$, computed using $\Sigma$ and $\PaVec^{\prime}$
%as the collection of parent sets of the intervention nodes.
Due to Theorem \ref{theorem: soundness of the semi-local algorithm}, $\Theta_p^{(1,\dots,k)}$ and $\Theta_p^{*(1,\dots,k)}$ (see \eqref{eq: Theta*}) are equivalent multisets.
%The multi-set of possible total effects of $X_i$ on $X_p$ in a joint intervention on $(X_1,\dots, X_k)$ is defined as follows and can easily be obtained from $\Theta_p^{(1,\ldots,k)}$:\vspace{-0.1in}
%\begin{align}\label{eq: Theta ip}
%   \Theta_{ip}^{(1,\ldots,k)} := \{ \effect_{ip}^{(1,\ldots,k)}(\PaVec^{\prime}) \st \PaVec^{\prime} \in \mathcal{PA}_{s\ell} \}.
%\end{align}
%\vspace{-0.1in}

In Section \ref{A2-section: comparing OPIN methods} of \cite{NandyMaathuisRichardson14b}, we illustrate and compare the OPIN methods RRC, MCD or IPW \cite{RobinsHernanBrumback00} when the CPDAG is known. We define joint-IDA estimators of $\Theta_p^{(1,\dots,k)}$ in Section \ref{subsection: joint IDA estimator}, when the underlying CPDAG is not known, by combining a structure learning algorithm, such as the PC-algorithm \cite{SpirtesEtAl00,ColomboEtAl14}, with an OPIN method. Consistency in sparse high-dimensional settings is proved in Section \ref{subsection: high-dim consistency} for the joint-IDA estimator based on RRC and MCD when the error variables are Gaussian and the PC algorithm is used for estimating the underlying CPDAG.

%%%%%%%%%%%%%%%%%%%%%%%%%%%%%%%%%%%%%%%%%%%%%%%%%%%%%%%%%%%%%%%%%%%%%%%%%%%%%%%%%%%%%%%
\subsection{Joint-IDA estimator}\label{subsection: joint IDA estimator}

Suppose that we have $n$ i.i.d.\ observations from a linear SEM characterized by $(\Dag,\errorvec)$, where $\Dag$ is unknown. Then we can estimate $\Theta_p^{(1,\dots,k)}$ using Algorithm \ref{algorithm: estimation from observational data}:
\begin{algorithm}[H]
   \caption{Joint-IDA estimator}
   \label{algorithm: estimation from observational data}
   \begin{algorithmic}[1]
   \REQUIRE $n$ i.i.d.\ observations of $\allvar$ (data), intervention nodes $(1,\ldots, k)$, OPIN method
   \ENSURE Estimate of $\Theta_p^{(1,\dots,k)}$ (see \eqref{eq: Theta})
   \STATE obtain an estimate of the underlying CPDAG $\hat{\mathcal{C}}$ from observational data (e.g., using the PC-algorithm);
   %\STATE $\hat{\mathcal{C}}(\alpha) \leftarrow$ output of the PC-algorithm when applied to data with tuning parameter $\alpha$;
   \STATE $\widehat{\mathcal{PA}}_{s\ell} \leftarrow$ output of the sample version of Algorithm \ref{algorithm: semi-local algorithm} when applied to $\hat{\mathcal{C}}$;
   \STATE for each $\PaVec^{\prime} \in \widehat{\mathcal{PA}}_{s\ell} $, let $\bar{\boldsymbol\theta}_p^{(1,\dots,k)}(\PaVec^{\prime})$ be an estimate of $\boldsymbol\theta_p^{(1,\dots,k)}$ using parent sets $\PaVec^{\prime}$ and the given OPIN method;\label{step: method}
   \RETURN $\bar{\Theta}_p^{(1,\dots,k)} = \{\bar{\boldsymbol\theta}_{p}^{(1,\ldots,k)}(\PaVec^{\prime}) \st \PaVec^{\prime} \in \widehat{\mathcal{PA}}_{s\ell} \}$.
   \end{algorithmic}
\end{algorithm}

In Section \ref{A2-section: low-dimensional simulation} of \cite{NandyMaathuisRichardson14b}, we compare the performance of the joint-IDA estimators based on RRC, MCD or IPW in simulation studies with a low dimensional setting.

% \textcolor{red}{In Section \ref{A2-section: low-dimensional simulation} of \cite{NandyMaathuisRichardson14b}, we compare joint-IDA estimators in simulation settings where the generated DAGs are comparable with the DAG in Example \ref{ex: simulation example}. To this end, we apply Algorithm \ref{algorithm: estimation from observational data}, using true CPDAGs and estimated CPDAGs separately in step 1 of the algorithm, and using RRC, MCD and IPW in step 3 of the algorithm. Simulation results show that RRC and MCD perform about equally well and outperform IPW. The performance of IPW is closer to that of RRC and MCD when the CPDAG is estimated. This can be explained by the fact that the error coming from the estimation of the CPDAG is significant and common to all three methods.}

%%%%%%%%%%%%%%%%%%%%%%%%%%%%%%%%%%%%%%%%%%%%%%%%%%%%%%%%%%%%%%%%%%%%%%%%%%%%%%%%%%%%%%%
\subsection{High-dimensional consistency with Gaussian errors}\label{subsection: high-dim consistency}

We now consider consistency of our methods, using an asymptotic scenario where the causal DAGs and the number of variables are allowed to change with $n$. Thus, let $\Dag_n = (\VSet_n,\ESet_n)$ and $\errorvec_n$ be sequences of causal DAGs and Gaussian error vectors, where $\VSet_n = \{ 1,\ldots, p_n \}$ and $\errorvec_n = (\error_{n1},\ldots,\error_{np_n})^T$. Let $\allvar_n := ( \rvX_{n1},\ldots,\rvX_{np_n} )^T$ be generated from the Gaussian linear SEM characterized by $(\Dag_n,\errorvec_n)$. Moreover, assume that we have $n$ i.i.d.\ observations from the multivariate Gaussian distribution of $\allvar_n$, for all $n$.

Consistency of the IDA algorithm in sparse high dimensional settings was shown under the following assumptions \cite{MaathuisKalischBuehlmann09}:
\begin{description}
   \item[(A1)] (Gaussianity and faithfulness) The distribution $\allvar_n $ is Gaussian and faithful to the true underlying DAG $\Dag_n$ for all $n$;
   \item[(A2)] (high-dimensional setting) $p_n = \mathcal{O}(n^{a})$ for some $0\leq a < \infty$;
   \item[(A3)] \label{assumption 3} (sparsity condition) Let $q_n = \max_{1 \leq i \leq p_n} |\NeSet_{i}(\Dag_n)|$.
        %be the maximum neighbourhood size in $\Dag_n$, where $\NeSet_{i}(\Dag_n)$ denotes the adjacency set of node $i$ in $\Dag_n$.
        Then $q_n = \mathcal{O}(n^{1-b})$ for some $0 < b \leq 1$;
   \item[(A4)] (bounds on partial correlations) The partial correlations $\rho_{nij | S}$ between $\rvX_{ni}$ and $\rvX_{nj}$ given $\{\rvX_{nr} : r \in S\}$ satisfy the following upper and lower bounds for all $n$,
       %uniformly over $i,j \in \{1,\ldots,p_n \}$ and $S \subseteq \{1,\ldots,p_n\}\setminus \{ i,j\}$ such that $|S|\leq q_n$:\vspace{-0.05in}
   \begin{align*}
    \mathop{\sup}_{i\neq j, |S|\le q_n} |\rho_{nij | S}| \leq M < 1,  ~~ \text{and} ~~  \mathop{\inf}_{i, j, |S|\le q_n} \{|\rho_{nij | S}| \st \rho_{nij | S} \neq 0 \} \geq c_n,\vspace{-0.05in}
   \end{align*}
   with $c_n^{-1} = \mathcal{O}(n^d)$ for some $0 < d < b/2$ where $0 < b \leq 1$ is as in (A3).
   \item[(A5)] Let $\{\Dag_{n1},\ldots,\Dag_{nm_n} \}$ be the Markov equivalence class of $\Dag_n$. Then for some $v>0$, \vspace{-0.1in}
   \begin{align*}
   \mathop{\sup}_{ i < p_n,r\leq m_n} \frac{\Var (\rvX_{np_n} | \rvX_{ni},  \PaVec_i(\Dag_{nr}))}{\Var (\rvX_{ni} | \PaVec_{i}(\Dag_{nr}))} \leq v.
   \end{align*}
\end{description}
Assumptions (A1)-(A4) are required for consistency of the PC-algorithm \cite{KalischBuehlmann07a}. We note that assumption (A5) is slightly weaker than the assumption made by the authors in \cite{MaathuisKalischBuehlmann09}, where for each $i$ they took the supremum over all possible subsets of $\NeSet_i(\Dag_n)$ instead of $\{\PaVec_i(\Dag_{nr}) \st r \leq m_n\}$. However, (A5) is sufficient for the proof presented in \cite{MaathuisKalischBuehlmann09}.

We will now show a similar consistency result for the joint-IDA estimator, based on either RRC or MCD. For simplicity, we only consider double interventions.
In particular, we consider all multisets of total joint effects of $\rvX_{ni}$ and $\rvX_{nj}$ on $\rvX_{np_n}$, defined as \begin{align}
   \Theta_{p_n}^{(i,j)} = \{ \effectvecn^{(i,j)}(\PaVec^{\prime})=(\effect_{ip_n}^{(i,j)}(\PaVec'), \effect_{jp_n}^{(i,j)}(\PaVec'))^T \st \PaVec^{\prime} \in \mathcal{PA}_{n,s\ell}^{(i,j)} \},
\end{align}
for $i \neq j \in \{1,\ldots,p_n-1\}$, where $\mathcal{PA}_{n,s\ell}^{(i,j)}$ is the output of Algorithm \ref{algorithm: semi-local algorithm} for intervention nodes $(i,j)$.

The output of the PC algorithm depends on a tuning parameter $\alpha_n$ and thus we denote the corresponding joint-IDA estimators based on RRC and MCD by $\hat \Theta_{p_n}^{(i,j)}(\alpha_n)$ and $\tilde \Theta_{p_n}^{(i,j)}(\alpha_n)$, respectively (see Algorithm \ref{algorithm: estimation from observational data}).
 %where, as before, we use a hat to denote the estimator based on RRC and a tilde for the estimator based on MCD.
 Our goal is to show that distance between $\Theta_{p_n}^{(i,j)}$ and $\hat{\Theta}_{p_n}^{(i,j)}(\alpha_n)$ (or $\tilde{\Theta}_{p_n}^{(i,j)}(\alpha_n)$) converges to zero in probability, uniformly over $i$ and $j$, under some suitable distance measure. To this end, we define the following distance between multisets of 2-dimensional vectors.

\begin{definition}\label{definition: distance measure} (Distance between multisets)
For any two multisets of scalars $A = \{ a_1,\ldots,a_{m_1} \}$ and $B = \{b_1,\ldots,b_{m_2}\}$ with order statistics $a_{(1)},\ldots,a_{(m_1)}$ and $b_{(1)},\ldots,b_{(m_2)}$, we define
\begin{align*}
{d}_{}(A,B) = \left\{ \begin{array}{ll} \mathop{\sup}_{i=1,\ldots,m_1} |a_{(i)} - b_{(i)} | & \text{if $m_1 = m_2$} \\
														\infty & \text{if $m_1 \neq m_2$} \end{array} \right. .
\end{align*}
For multisets of 2-dimensional vectors $A = \{ (a_{11},a_{21})^T,\ldots,(a_{1m_1},a_{2m_1})^T\}$ and $B = \{ (b_{11},b_{21})^T,\ldots,(b_{1m_2},b_{2m_2})^T\}$, we define
\begin{align*}
{d}_{}(A,B) = \max({d}_{}(A_1,B_1),{d}_{}(A_2,B_2)),
\end{align*}
where $A_i = \{ a_{i1},\ldots,a_{im_1} \}$ and $B_i = \{b_{i1},\ldots,b_{im_2}\}$, for $i=1,2$.
\end{definition}

We consider the following modifications of assumption (A5):
\begin{itemize}
   \item[(A$5^*$)] Let $\{\Dag_{n1},\ldots,\Dag_{nm_n} \}$ be the Markov equivalence class of $\Dag_n$. Then for some $v^*>0$,\vspace{-0.05in}
   \begin{align*}
      \mathop{\sup}_{\substack{i <p_n ,j \leq p_n, r\leq m_n }} \frac{\Var (\rvX_{nj} | \PaVec_{i}(\Dag_{nr}))}{\Var (\rvX_{ni} | \PaVec_{i}(\Dag_{nr}))}  \leq v^{*}.
   \end{align*}
   \item[(A$5'$)] Let $\{\Dag_{n1},\ldots,\Dag_{nm_n} \}$ be the Markov equivalence class of $\Dag_n$, and let $\covmat_{nijr}$ and $\covmat_{nijr}^{\prime}$ denote the covariance matrices of  $\mathbf{U}_{nijr} := \{ \rvX_{np_n}\} \cup \{\rvX_{ni},\PaVec_{i}(\Dag_{nr}) \} \cup \{\rvX_{nj},\PaVec_{j}(\Dag_{nr}) \}$ and $\mathbf{U}_{nijr} \setminus \{\rvX_{np_n}\}$, respectively. Then for some $v^{\prime}>0$,\vspace{-0.05in}
   \begin{align*}
      \mathop{\sup}_{i<p_n,j<p_n, r \leq m_n} ||\covmat_{nijr}^{\prime -1}||~ ||\covmat_{nijr}|| = \mathop{\sup}_{i<p_n,j<p_n, r \leq m_n} \frac{ \lambda_{\max} (\covmat_{nijr})}{\lambda_{\min} (\covmat_{nijr}^{\prime})} \leq v^{\prime},
   \end{align*}
   where for any matrix $A$, $\lambda_{\max}(A)$ (or $\lambda_{\min}(A)$) is the maximum (or minimum) eigenvalue of $A$ and $||A|| = \sqrt{\lambda_{\max}(A^TA)}$ represents the spectral norm.
\end{itemize}

Assumption (A5') is stronger than (A5*), and (A5*) is stronger than (A5) (see Section \ref{A2-section: comparison of assumptions} of \cite{NandyMaathuisRichardson14b} for a detailed discussion). We now obtain the following consistency results.
\begin{theorem}\label{theorem: asymptotic consistency of RRC} (Consistency of RRC)
   Under assumptions (A1)-(A4) and (A$5^*$), there exists a sequence $\alpha_n$ converging to zero such that
   \begin{align*}
      \mathop{\sup}_{i<p_n,j<p_n, i \neq j} {d}_{}(\hat{\Theta}_{p_n}^{(i,j)}(\alpha_n),\Theta_{p_n}^{(i,j)}) \stackrel{\Prob}{\longrightarrow} 0.
   \end{align*}
\end{theorem}

\begin{theorem}\label{theorem: asymptotic consistency of MCD} (Consistency of MCD)
   Under assumptions (A1)-(A4) and (A$5'$), there exists a sequence $\alpha_n$ converging to zero such that
   \begin{align*}
   \mathop{\sup}_{i<p_n,j<p_n, i\neq j} {d}_{}(\tilde{\Theta}_{p_n}^{(i,j)}(\alpha_n),\Theta_{p_n}^{(i,j)}) \stackrel{\Prob}{\longrightarrow} 0.
   \end{align*}
\end{theorem}

We define the multi-set of possible total effects of $X_{ni}$ on $X_{np_n}$ in a joint intervention on $(X_{ni},X_{nj})$  as \begin{align*}
   \Theta_{ip_n}^{(i,j)} := \{ \effect_{ip_n}^{(i,j)}(\PaVec^{\prime}) \st \PaVec^{\prime} \in \mathcal{PA}_{n,s\ell}^{(i,j)} \}.
\end{align*}
%\begin{align}\label{eq: Theta ip}
%   \Theta_{ip}^{(i,j)} := \{ \effect_{ip}^{(i,j)}(\PaVec^{\prime}) \st \PaVec^{\prime} \in \mathcal{PA}_{s\ell}^{(i,j)} \}.
%\end{align}
Let $\hat{\Theta}_{ip_n}^{(i,j)}(\alpha_n)$ and $\tilde{\Theta}_{ip_n}^{(i,j)}(\alpha_n)$ be the corresponding joint-IDA estimators using RRC and MCD, respectively. Then Definition \ref{definition: distance measure} and Theorems \ref{theorem: asymptotic consistency of RRC} and  \ref{theorem: asymptotic consistency of MCD} guarantee that certain summary measures of the estimated multisets converge in probability to the corresponding summary measures of ${\Theta}_{ip_n}^{(i,j)}$, uniformly over $i$ and $j$. We state this result in a corollary.
\begin{corollary} (Consistency of summary measures)\label{cor: consistency summary measures}
   Under the assumptions of Theorem \ref{theorem: asymptotic consistency of RRC} (for RRC) or Theorem \ref{theorem: asymptotic consistency of MCD} (for MCD), there exists a sequence $\alpha_n$ converging to zero such that the following sequences converge to zero in probability:
   \begin{enumerate}
      \item $\mathop{\sup}_{i<p_n,j<p_n, i \neq j} |\mathrm{minabs}(\bar{\Theta}_{ip_n}^{(i,j)}(\alpha_n)) - \mathrm{minabs}({\Theta}_{ip_n}^{(i,j)}) |$,
      \item $\mathop{\sup}_{i<p_n,j<p_n, i \neq j} |\mathrm{aver}(\bar{\Theta}_{ip_n}^{(i,j)}(\alpha_n)) - \mathrm{aver}({\Theta}_{ip_n}^{(i,j)}) | $,
   \end{enumerate}
   where $\bar{\Theta}_{ip_n}^{(i,j)}(\alpha_n)$ denotes  $\hat{\Theta}_{ip_n}^{(i,j)}(\alpha_n)$ or $\tilde{\Theta}_{ip_n}^{(i,j)}(\alpha_n)$, and $\mathrm{minabs}(A) := \min\{|a| : a \in A \}$ and $\mathrm{aver}(A) := |A|^{-1} \sum_{a \in A} a$.
%
%      \item $\mathop{\sup}_{i<p_n,j<p_n, i \neq j} |\mathrm{minabs}(\tilde{\Theta}_{ip_n}^{(i,j)}(\alpha_n)) - \mathrm{minabs}({\Theta}_{ip_n}^{(i,j)}) |$,
%      \item $\mathop{\sup}_{i<p_n,j<p_n, i \neq j} |\mathrm{aver}(\tilde{\Theta}_{ip_n}^{(i,j)}(\alpha_n)) - \mathrm{aver}({\Theta}_{ip_n}^{(i,j)}) |$,
%   \end{enumerate}
\end{corollary}

In Section \ref{A2-section: high-dimensional simulation} of \cite{NandyMaathuisRichardson14b}, we show high-dimensional simulation studies that provide empirical support for
 Corollary \ref{cor: consistency summary measures}. Using these simulations to compare the performances of the joint-IDA estimators based on RRC, MCD and IPW, we find that RRC and MCD outperform IPW. In Section \ref{A2-section: application} of \cite{NandyMaathuisRichardson14b}, we apply the joint-IDA estimators to gene expression data from the DREAM4 challenge \cite{SchaffterEtAl11,MarbachEtAl10} with two specific goals. The first goal is to identify triples of genes for which the total effect of simultaneous knock-out of the first two genes on the third gene are large, and the second goal is to identify triples of genes for which the strength of the epistatic interaction \citep{JasnosKorona07, VelenichGore13} between the first two genes for regulating the third gene is high. In both cases, the joint-IDA estimators perform reasonably well and much better than random guessing, showing the usefulness of the joint-IDA estimators in such applications. In Section \ref{A2-section: validation on simulated data} of \cite{NandyMaathuisRichardson14b}, we perform further simulation studies in high-dimensional settings with a mixture of Gaussian and non-Gaussian errors, where, motivated by the application on DREAM4 data, we aim to identify intervention sets and response variables for which the total effects of joint interventions are large. As in the high-dimensional simulations in Section \ref{A2-section: high-dimensional simulation} of \cite{NandyMaathuisRichardson14b}, we again find that the joint-IDA estimators based on RRC and MCD outperform the joint-IDA estimator based on IPW.

%% file: texfiles/NPN-jointIDA.tex
\section{A relaxation of the linearity assumption}\label{section: relaxing linearity}
So far, we assumed that the data are generated from a \emph{linear} SEM with independent continuous errors.
%The linearity assumption that requires that each node is a linear function of its parents and random noise, is key for our methods for estimating the total effects in a joint intervention. Dropping the linearity assumption would fundamentally change the problem. This is mainly because without this assumption, the total effects would generally be functions (of the values fixed by interventions) instead of scalars, and such functions are much harder to estimate.
In this section, we show that a simple modification of the joint-IDA estimators can be applied to nonparanormal distributions \citep{LiuEtAl09, HarrisDrton13}, which form an interesting non-linear and non-Gaussian generalization of linear Gaussian SEMs.

\begin{definition}\label{definition: nonparanormal} (Nonparanormal distribution \citep{HarrisDrton13})
Let $\mathbf{g} = ( g_1,\ldots,g_{p})^T$ be a collection of strictly increasing functions $g_i : \mathbb{R} \to \mathbb{R}$, and let $\Sigma_0$ be a positive definite correlation matrix. The nonparanormal distribution $NPN(\mathbf{g} , \Sigma_0)$ is the distribution of the random vector $( g_1(Z_1),\ldots,g_{p}(Z_p))^T$ for $\mathbf{Z} = (Z_1,\ldots,Z_p)^T \sim N(0, \Sigma_0)$.
\end{definition}

Let $\mathbf g$ and $\mathbf Z$ be as in Definition \ref{definition: nonparanormal} and let $\mathbf{X} = ( g_1(Z_1),\ldots,g_{p}(Z_p))^T$. Assuming an underlying structural equation model in terms of $\mathbf{Z}$:
\begin{equation}\label{eq:linear-sem-z}
   \mathbf Z \leftarrow B^T{\bf Z} + {\boldsymbol \epsilon},
\end{equation}
we obtain a non-linear and non-Gaussian structural equation model in terms of $\mathbf{X}$:
\begin{equation} \label{eq:npn-sem}
   \mathbf{X}  \leftarrow \mathbf{g}(B^T\mathbf{g}^{-1}({\bf X}) + {\boldsymbol \epsilon}),
\end{equation}
where $\mathbf{g}^{-1}(\cdot) \equiv (g_1^{-1}(\cdot),\ldots ,g_p^{-1}(\cdot))$ is the map taking $(X_1,\ldots , X_p)$ to $(Z_1,\ldots , Z_p)$.
%We emphasize here  that we treat the functions $g_i$ as purely definitional (not structural).

There is a one-to-one correspondence between interventions on the $Z$'s in the linear system \eqref{eq:linear-sem-z} and interventions on the $X$'s in the non-linear system \eqref{eq:npn-sem}: if intervening to set $(Z_1, \dots ,Z_k)$ to  $(z_1,\dots, z_k)$ in (\ref{eq:linear-sem-z}) results in $Z_p$ taking the value $z_p$, then setting $(X_1, \dots ,X_k)$ to  $(x_1 = g_1(z_1),\dots, x_k = g_k(z_k))$ in \eqref{eq:npn-sem} will result in $X_p$ taking the value $x_p = g_p(z_p)$. The total joint effect of $(X_1,\dots,X_k)$ on $X_p$ in \eqref{eq:npn-sem} is difficult to summarize due to the nonlinearities. However, non-zero total joint effects among the $Z$'s correspond to non-zero total joint effects among the $X$'s (and vice-versa). In the remainder, we therefore focus on estimating total joint effects among the $Z$'s based on observational data from the distribution of $\mathbf{X}$.

Since the $g_i$'s are increasing functions, the (sample) rank correlation coefficient (Spearman's $\rho$ or Kendall's $\tau$) between $X_i$ and $X_j$ and between $Z_i$ and $Z_j$ are identical. Further, \cite{LiuEtAl12} showed that excellent estimators of the Pearson correlation coefficient between two jointly Gaussian random variables can be obtained by taking trigonometric transformations of their sample rank correlation coefficients. In particular, if $(Z_i,Z_j)$ are bivariate normal with Pearson correlation coefficient $\rho_{ij}$, then
 \begin{align*}
 &\Prob (| 2\sin(\frac{\pi}{6} \hat{\rho}^{S}_{ij})  - \rho_{ij}| > \epsilon) \leq 2 \exp (-\frac{2}{9\pi^2}n\epsilon^2),~\text{and} \\
 &\Prob (| \sin(\frac{\pi}{2} \hat{\rho}^{K}_{ij})  - \rho_{ij}| > \epsilon) \leq 2 \exp (-\frac{2}{\pi^2}n\epsilon^2),
 \end{align*}
where $\hat{\rho}^{S}_{ij}$ and $\hat{\rho}^{K}_{ij}$ denote the sample Spearman's rank and Kendall's rank correlations based on $n$ i.i.d.\ data.

Thus, we propose to use RRC or MCD with an estimate of $\Sigma_0$ given by $(\hat{\Sigma}_0)_{ij} = 2\sin(\frac{\pi}{6} \hat{\rho}^{S}_{ij})$ or $(\hat{\Sigma}_0)_{ij} = \sin(\frac{\pi}{2} \hat{\rho}^{K}_{ij})$ when the parent sets of the intervention variables are given, for estimating the total joint effect of $(Z_1,\ldots,Z_k)$ on $Z_p$ based on i.i.d.\ data from the distribution of $\mathbf{X}$. We refer to these OPIN methods as NPN-RRC and NPN-MCD. Combining these methods with the Rank PC algorithm  \cite{HarrisDrton13}, we obtain NPN-joint-IDA estimators.

\begin{definition} (NPN-joint-IDA estimators)
The output of Algorithm \ref{algorithm: estimation from observational data} is an NPN-joint-IDA estimator of the total joint effect of $(Z_1,\ldots,Z_k)$ on $Z_p$ based on i.i.d.\ data from the distribution of $\mathbf{X}=( g_1(Z_1),\ldots,g_{p}(Z_p))$, when the Rank PC algorithm of \cite{HarrisDrton13} is used for estimating the CPDAG and NPN-RRC or NPN-MCD is used as OPIN method.
\end{definition}

A somewhat related method, called NPN-IDA, has been proposed by \cite{TeramotoEtAl14} for single interventions. However, in that work, the nonparanormal distribution appears to be used solely to estimate the CPDAG, with linear regression still used to estimate effects. Moreover, we provide theoretical guarantees NPN-joint-IDA estimators, such as the following high-dimensional consistency result:

%However, they only seem to use the nonparanormal distribution to estimate the CPDAG, and perform regular regressions afterward. In contrast, we use the nonparanormal distribution in both the CPDAG estimation and the OPIN estimation. Moreover, we provide theoretical guarantees for our method.

%When applying NPN-joint-IDA in practice, $\hat{\Sigma}$ (as defined via $\hat{\Sigma}_{ij} = 2\sin(\frac{\pi}{6} \hat{\rho}^{S}_{ij})$ or $\hat{\Sigma}_{ij} = \sin(\frac{\pi}{2} \hat{\rho}^{K}_{ij})$) may not be positive semidefinite. In such cases we slightly modify $\hat{\Sigma}$ by replacing negative eigenvalues of $\hat{\Sigma}$ by zeroes in its spectral decomposition.

%We now present a high-dimensional consistency result for the NPN-joint-IDA estimators.

\begin{theorem}\label{theorem: NPN high dimensional consistency}
Let $\mathbf{Z}_n = (Z_{n1},\ldots,Z_{np_n})^T \sim N(0, \Sigma_{n0})$, where $\Sigma_{n0}$ is a positive definite correlation matrix. Assume that the distribution of $\mathbf{X}_n = ( g_{n1}(Z_{n1}),\ldots,g_{np}(Z_{np_n}))^T$ is $NPN(\mathbf{g}_n , \Sigma_{n0})$. Further, assume (A2)-(A4) from Section \ref{section: estimation from observational data} for $\mathbf{Z}_n$, but with constants $b,d$ satisfying $2/3<b\leq 1$ and $0\leq d<b-1/2$. Moreover, assume (A$5^\prime$) from Section \ref{section: estimation from observational data} for $\mathbf{Z}_n$. Then there exists a sequence of $\alpha_n$ converging to zero such that
\begin{align*}
      \mathop{\sup}_{i<p_n,j<p_n, i \neq j} {d}_{}(\bar{\Theta}_{p_n}^{(i,j)}(\alpha_n),\Theta_{p_n}^{(i,j)}) \stackrel{\Prob}{\longrightarrow} 0,
   \end{align*}
where $d(\cdot,\cdot)$ is given by Definition \ref{definition: distance measure}, $\Theta_{p_n}^{(i,j)}$ is the multi-set of possible total joint effects of $(Z_{n1},\dots,Z_{nk})$ on $Z_{np_n}$, and $\bar{\Theta}_{p_n}^{(i,j)}(\alpha_n)$ denotes the corresponding NPN-joint-IDA estimators based on RRC or MCD and the Rank PC algorithm with tuning parameter $\alpha_n$.
\end{theorem}

 In Section \ref{A2-section: validation on simulated data 2} of \cite{NandyMaathuisRichardson14b}, we investigate the performances of the NPN-joint-IDA estimators in high-dimensional settings with nonparanormal distributions and also under a slight violation of the nonparanormal assumption, namely when $\mathbf{Z}$ is generated from a linear SEM with a mixture of Gaussian and non-Gaussian errors and $\allvar$ is a monotone transformation of $\mathbf{Z}$ (as in Definition \ref{definition: nonparanormal}). The NPN-joint-IDA estimators perform about equally well in these two cases, suggesting  insensitivity of the NPN-joint-IDA estimators to such slight violations of the nonparanormal assumption.

%% file: texfiles/discussion.tex
\section{Discussion}\label{section: discussion}
%\textbf{(please ignore this part)}\\\\
%We consider for the first time the problem of estimating the effect of multiple interventions from observational data, when assuming that the true causal DAG is not known. We propose three methods for this problem: IDA-path is straightforward. IDA-IPW is a new combination of the PC-algorithm and IPW, exploiting a local algorithm which is similar to the one used in IDA. The methods based on modifying the Cholesky decomposition are novel.\\
%TODO: Intro... Problem we consider, why interesting, main results - design of experiments. Two very different methods
We considered the problem of estimating the effect of multiple simultaneous interventions, based on observational data from an \emph{unknown} linear SEM with continuous errors, or equivalently, from an \emph{unknown} linear DAG model, in sparse high-dimensional settings. There is previous work on estimating causal effects of single interventions from unknown Gaussian DAGs in high-dimensional settings (e.g., \cite{MaathuisKalischBuehlmann09}), as well as work on estimating the effect of multiple simultaneous interventions from observational data when the underlying causal DAG is given (e.g., \cite{RobinsHernanBrumback00}), but considering the combination of these different aspects seems to be novel. Thus, we provide a first approach to address this problem, including theoretical guarantees as well as evaluations on simulated and in-silico data.

As a stepping stone, we first considered a scenario where we have partial knowledge of the underlying causal DAG, in the sense that we know only the parents of the intervention nodes (OPIN assumption). We introduced two new methods for estimating total joint effects in this setting, called RRC and MCD. Both methods are based on original ideas. RRC uses a novel recursive relation to determine the total joint effect of a multiple intervention of arbitrary cardinality from single intervention effects. MCD is based on several modified Cholesky decompositions of the covariance matrix, where the given parent information is used to re-order the variables appropriately in each Cholesky decomposition. We note that we do not need to use the full covariance matrix of $\allvar$, but only the low-dimensional sub-matrix corresponding to the intervention nodes, their parents and the variable of interest. We showed in simulations that RRC and MCD typically outperform IPW \cite{RobinsHernanBrumback00} in the Gaussian setting. The general question of efficiency under the OPIN assumption, however, is open. It would be very interesting to determine optimally efficient estimators under the OPIN assumption in parametric, semi-parametric or non-parametric settings.

Next, we defined a joint-IDA estimator (Algorithm \ref{algorithm: estimation from observational data}) for estimating multisets of possible total joint effects from observational data from an unknown linear SEM. The joint-IDA estimator consists of three steps: (1) estimating the CPDAG of the underlying causal DAG, (2) extracting possible jointly valid parent sets of the intervention nodes from the CPDAG (Algorithm \ref{algorithm: semi-local algorithm}), and (3) an OPIN method from Section \ref{section: multiple interventions using parent sets}. This combination of methods was chosen because it scales well to large sparse graphs. In step (2) we use a semi-local algorithm that preserves multiplicity information at a low computational cost. This algorithm can also be used in IDA for single interventions when multiplicity information is important. The use of OPIN methods in step (3) ensures that we only require semi-local information of the CPDAG around the intervention nodes, making the method insensitive to estimation errors in the CPDAG that occur ``far away" from the intervention nodes.
%
%By combining the OPIN methods with the PC-algorithm and a method to extract possible
%We consider the estimation of the total joint effect from observational data generated from a Gaussian linear SEM. A motivating example is high-dimensional gene expression data where we aim to predict the effect of double or triple gene knockout on other genes or some phenotype of interest. The gold-standard method for estimating such effects is gene knockout experiments. However, finding all significant double or triple gene knockout effects by the gold-standard method requires a huge number of experiments and a tool for estimating (bounds on) joint effects from observational data can help to reduce the number of experiments significantly. We propose two new methods, RRC and MCD, for estimating the total joint effect when the parent sets of the intervention variables in the causal DAG are known and combine them with the PC-algorithm to accomplish our goal.
%

We derived the asymptotic distributions of the RRC and MCD estimators under the OPIN assumption. Moreover, we proved consistency of the joint-IDA estimator based on RRC or MCD in sparse high-dimensional settings with Gaussian noise. These analyses were rather non-standard for the MCD estimator, due to the special nature of this algorithm.

In simulations, the joint-IDA estimators based on RRC or MCD outperformed the one based on IPW, where RRC might have a small advantage over MCD. MCD is more general, however, in the sense that it not only yields the total joint effect, but also the post-intervention covariance matrix. The total joint effect is one quantity that can be computed from this matrix, but other quantities may be of interest as well. Moreover, the joint-IDA estimator based on MCD can be easily generalized to settings with so-called mechanism changes \cite{Tian01}, where one wants to know what happens if a node depends on (a subset of) its parents in a different way, in the sense that the edge weights in the linear SEM are changed. (Pearl's do-operator can be viewed as a special case of a mechanism change, where the intervention node no longer depends on its parents at all.) For example, one may want to predict the effect of a change of policy (e.g. tax reform, labor dispute resolution) in an economic model. In biochemistry, it may be interesting to know what happens to a biochemical network (e.g. gene regulatory network, protein-protein network) if one blocks one of the two or three binding sites of some biochemical agents (e.g. gene, peptides). Activity interventions as considered in \cite{MooijHeskes13} can also be represented by mechanism changes. In MCD, such mechanism changes can be incorporated by simply setting the entries in the Cholesky decomposition to the negative new edge weights, rather than to zero. Mechanism changes can also be easily incorporated in IPW by modifying the weights, while there seems no straightforward modification for RRC.

%\textcolor{red}{We pointed out that in some sparse networks most of the joint effects (or single intervention effects) can be equal (or close) to the corresponding unadjusted regression coefficients. In such a case, (joint)-IDA-like method, in general, can bring no significant improvement over the naive method of regression without any covariate adjustment (or estimating causal effects by correlation). However, we showed with some examples that joint-IDA can improve estimation in some realistic settings, where we ensure to have some hubs in the network. Note that this comparison is not new. IDA has been validated on real datasets where it has performed significantly better than the marginal correlation screening \cite{StekhovenEtAl12}.}

%Another direction for generalization is to relax the Gaussian assumption. Under the assumption that the data are generated from a linear SEM where the errors are non-Gaussian, LiNGAM based methods \cite{ShimizuEtAl06-JMLR, ShimizuEtAl11} can be used to estimate the underlying DAG. Then RRC or MCD can be applied to estimate the total joint effect as they do not depend on the Gaussian assumption.
In Sections \ref{section: introduction} - \ref{section: estimation from observational data} we assumed an underlying linear SEM with independent continuous errors. %mainly assume that there are no hidden confounders and each variable is a linear function of its parents.
In practice, both the linearity and the independence of the errors (absence of hidden confounders) can be violated. Section \ref{section: relaxing linearity} discussed a generalization to nonparanormal distributions, allowing for non-linearity. As a possible direction for future work, one may also try to relax the no hidden confounders assumption, for example by combining a method to estimate the Markov equivalence class of so-called maximal ancestral graphs (MAGs) (see, e.g., \cite{SpirtesEtAl00, ColomboEtAl12, ClaassenMooijHeskes13}) and the adjustment criteria developed in \cite{MaathuisColombo14, vanderZarderEtAl14, PerkovicEtAl15}.

%The linearity assumption requires that each node is a linear function of its parents and random noise. This assumption is key for our methods. It also seems rather natural in the high-dimensional settings we consider, in the sense that fitting highly complicated models is infeasible in such scenarios. Dropping the linearity assumption would fundamentally change the problem. For example, without this assumption, causal effects would generally be functions (of the values fixed by interventions) instead of scalars, and such functions are much harder to estimate. Also, a non-parametric approach based on discretized data is infeasible, since we are not aware of any causal structure learning algorithms for discrete Bayesian networks that are consistent in high-dimensional settings. We note, however, that we do \emph{not} assume that every conditional expectation is linear, see Example \ref{A2-example: nonlinerity} in \cite{NandyMaathuisRichardson14b} for a concrete example. Moreover, when the parent set of a variable is known, the corresponding linearity assumption is actually testable from observational data, for example by a goodness of fit test of the linear model \cite{CookWeisberg82}.
%%In a recent paper, \cite{BuhlmannVandeGeer15} discussed robustness of linear models in high-dimensional inference under model misspecification.}

We emphasize that our methods should not be used as a replacement for randomized or interventional experiments. Rather, methods like (joint-)IDA can provide useful guidelines for prioritizing such experiments, especially in high-dimensional settings where there are many possible intervention experiments. Validations of IDA \citep{MaathuisColomboKalischBuehlmann10, StekhovenEtAl12} and of joint-IDA (see Section \ref{A2-section: application} of \cite{NandyMaathuisRichardson14b}) indicate that these methods can indeed be useful tools for the design of experiments, even when some of their assumptions are violated. Further, such randomized experiments provide a means to subject their assumptions to empirical tests.

%% file: texfiles/acknowledgements.tex
\section{Acknowledgements}

We thank the referees, the associate editor and the editor for their constructive comments, which have led to significant improvements in the paper. In particular, their comments have led us to relax the Gaussianity and linearity assumptions that were present in an earlier version of this paper.

%% file: Paper_aos.bbl
\begin{thebibliography}{39}
% BibTex style file: imsart-number.bst, 2013-01-28
% Default style options (sort=1,type=number).
% Used options (sort=1,type=number).

\bibitem{AliferisEtAl10}
\begin{barticle}[author]
\bauthor{\bsnm{Aliferis},~\bfnm{Constantin~F.}\binits{C.~F.}},
  \bauthor{\bsnm{Statnikov},~\bfnm{Alexander}\binits{A.}},
  \bauthor{\bsnm{Tsamardinos},~\bfnm{Ioannis}\binits{I.}},
  \bauthor{\bsnm{Mani},~\bfnm{Subramani}\binits{S.}} \AND
  \bauthor{\bsnm{Koutsoukos},~\bfnm{Xenofon~D.}\binits{X.~D.}}
(\byear{2010}).
\btitle{Local Causal and Markov Blanket Induction for Causal Discovery and
  Feature Selection for Classification {P}art {I}: Algorithms and Empirical
  Evaluation}.
\bjournal{J. Mach. Learn. Res.}
\bvolume{11}
\bpages{171--234}.
\end{barticle}
\endbibitem

\bibitem{BalkePearl94}
\begin{binproceedings}[author]
\bauthor{\bsnm{Balke},~\bfnm{A.~A.}\binits{A.~A.}} \AND
  \bauthor{\bsnm{Pearl},~\bfnm{J.}\binits{J.}}
(\byear{1994}).
\btitle{Probabilistic evaluation of counterfactual queries}.
In \bbooktitle{AAAI 1994}.
\end{binproceedings}
\endbibitem

\bibitem{Castelo06}
\begin{barticle}[author]
\bauthor{\bsnm{Castelo},~\bfnm{Robert}\binits{R.}} \AND
  \bauthor{\bsnm{Roverato},~\bfnm{Alberto}\binits{A.}}
(\byear{2006}).
\btitle{A robust procedure for gaussian graphical model search from microarray
  data with $p$ larger than $n$}.
\bjournal{J. Mach. Learn. Res.}
\bvolume{7}
\bpages{2621-2650}.
\end{barticle}
\endbibitem

\bibitem{ClaassenMooijHeskes13}
\begin{binproceedings}[author]
\bauthor{\bsnm{Claassen},~\bfnm{Tom}\binits{T.}},
  \bauthor{\bsnm{Mooij},~\bfnm{Joris}\binits{J.}} \AND
  \bauthor{\bsnm{Heskes},~\bfnm{Tom}\binits{T.}}
(\byear{2013}).
\btitle{Learning sparse causal models is not {NP}-hard}.
In \bbooktitle{UAI 2013}.
\end{binproceedings}
\endbibitem

\bibitem{Cochran38}
\begin{barticle}[author]
\bauthor{\bsnm{Cochran},~\bfnm{W.~G.}\binits{W.~G.}}
(\byear{1938}).
\btitle{The omission or addition of an independent variable in multiple linear
  regression}.
\bjournal{J. R. Statist. Soc. Suppl.}
\bvolume{5}
\bpages{171-176}.
\end{barticle}
\endbibitem

\bibitem{ColomboEtAl14}
\begin{barticle}[author]
\bauthor{\bsnm{Colombo},~\bfnm{D.}\binits{D.}} \AND
  \bauthor{\bsnm{Maathuis},~\bfnm{M.~H.}\binits{M.~H.}}
(\byear{2014}).
\btitle{Order-independent constraint-based causal structure learning}.
\bjournal{J. Mach. Learn. Res.}
\bvolume{15}
\bpages{3741-3782}.
\end{barticle}
\endbibitem

\bibitem{ColomboEtAl12}
\begin{barticle}[author]
\bauthor{\bsnm{Colombo},~\bfnm{D.}\binits{D.}},
  \bauthor{\bsnm{Maathuis},~\bfnm{M.~H.}\binits{M.~H.}},
  \bauthor{\bsnm{Kalisch},~\bfnm{M.}\binits{M.}} \AND
  \bauthor{\bsnm{Richardson},~\bfnm{T.~S.}\binits{T.~S.}}
(\byear{2012}).
\btitle{Learning high-dimensional directed acyclic graphs with latent and
  selection variables}.
\bjournal{Ann. Statist.}
\bvolume{40}
\bpages{294-321}.
\end{barticle}
\endbibitem

\bibitem{CoxWermuth04}
\begin{barticle}[author]
\bauthor{\bsnm{Cox},~\bfnm{D.~R.}\binits{D.~R.}} \AND
  \bauthor{\bsnm{Wermuth},~\bfnm{N.}\binits{N.}}
(\byear{2004}).
\btitle{Causality: a statistical view}.
\bjournal{Int. Statist. Rev.}
\bvolume{72}
\bpages{285-305}.
\end{barticle}
\endbibitem

\bibitem{CoxWermuth03}
\begin{barticle}[author]
\bauthor{\bsnm{Cox},~\bfnm{D.~R.}\binits{D.~R.}} \AND
  \bauthor{\bsnm{Wermuth},~\bfnm{N.}\binits{N.}}
(\byear{2003}).
\btitle{A general condition for avoiding effect reversal after
  marginalization}.
\bjournal{J. R. Statist. Soc. B}
\bvolume{65}
\bpages{937-941}.
\end{barticle}
\endbibitem

\bibitem{DorTarsi92}
\begin{btechreport}[author]
\bauthor{\bsnm{Dor},~\bfnm{Dorit}\binits{D.}} \AND
  \bauthor{\bsnm{Tarsi},~\bfnm{Michael}\binits{M.}}
(\byear{1992}).
\btitle{A simple algorithm to construct a consistent extension of a partially
  oriented graph}
\btype{Technical Report} No. \bnumber{R-185},
\bpublisher{Cognitive Systems Laboratory, UCLA}.
\end{btechreport}
\endbibitem

\bibitem{DrtonEtAl12}
\begin{barticle}[author]
\bauthor{\bsnm{Drton},~\bfnm{Mathias}\binits{M.}},
  \bauthor{\bsnm{Fox},~\bfnm{Chris}\binits{C.}} \AND
  \bauthor{\bsnm{K{\"a}ufl},~\bfnm{Andreas}\binits{A.}}
(\byear{2012}).
\btitle{Comments on: {S}equences of regressions and their independencies}.
\bjournal{TEST}
\bvolume{21}
\bpages{255--261}.
\end{barticle}
\endbibitem

\bibitem{HarrisDrton13}
\begin{barticle}[author]
\bauthor{\bsnm{Harris},~\bfnm{Naftali}\binits{N.}} \AND
  \bauthor{\bsnm{Drton},~\bfnm{Mathias}\binits{M.}}
(\byear{2013}).
\btitle{{PC} algorithm for nonparanormal graphical models}.
\bjournal{J. Mach. Learn. Res.}
\bvolume{14}
\bpages{3365-3383}.
\end{barticle}
\endbibitem

\bibitem{JasnosKorona07}
\begin{barticle}[author]
\bauthor{\bsnm{Jasnos},~\bfnm{Lukasz}\binits{L.}} \AND
  \bauthor{\bsnm{Korona},~\bfnm{Ryszard}\binits{R.}}
(\byear{2007}).
\btitle{Epistatic buffering of fitness loss in yeast double deletion strains}.
\bjournal{Nat. Genet.}
\bvolume{39}
\bpages{550 -- 554}.
\end{barticle}
\endbibitem

\bibitem{KalischBuehlmann07a}
\begin{barticle}[author]
\bauthor{\bsnm{Kalisch},~\bfnm{M.}\binits{M.}} \AND
  \bauthor{\bsnm{B\"uhlmann},~\bfnm{P.}\binits{P.}}
(\byear{2007}).
\btitle{Estimating high-dimensional directed acyclic graphs with the
  {PC}-algorithm}.
\bjournal{J. Mach. Learn. Res.}
\bvolume{8}
\bpages{613--636}.
\end{barticle}
\endbibitem

\bibitem{KalischEtAl12}
\begin{barticle}[author]
\bauthor{\bsnm{Kalisch},~\bfnm{M.}\binits{M.}},
  \bauthor{\bsnm{M\"achler},~\bfnm{M.}\binits{M.}},
  \bauthor{\bsnm{Colombo},~\bfnm{D.}\binits{D.}},
  \bauthor{\bsnm{Maathuis},~\bfnm{M.~H.}\binits{M.~H.}} \AND
  \bauthor{\bsnm{B\"uhlmann},~\bfnm{P.}\binits{P.}}
(\byear{2012}).
\btitle{Causal inference using graphical models with the {R} package pcalg}.
\bjournal{J. Statist. Software}
\bvolume{47}
\bpages{1-26}.
\end{barticle}
\endbibitem

\bibitem{LiuEtAl12}
\begin{barticle}[author]
\bauthor{\bsnm{Liu},~\bfnm{Han}\binits{H.}},
  \bauthor{\bsnm{Han},~\bfnm{Fang}\binits{F.}},
  \bauthor{\bsnm{Yuan},~\bfnm{Ming}\binits{M.}},
  \bauthor{\bsnm{Lafferty},~\bfnm{John}\binits{J.}} \AND
  \bauthor{\bsnm{Wasserman},~\bfnm{Larry}\binits{L.}}
(\byear{2012}).
\btitle{High-dimensional semiparametric Gaussian copula graphical models}.
\bjournal{Ann. Statist.}
\bvolume{40}
\bpages{2293--2326}.
\end{barticle}
\endbibitem

\bibitem{LiuEtAl09}
\begin{barticle}[author]
\bauthor{\bsnm{Liu},~\bfnm{Han}\binits{H.}},
  \bauthor{\bsnm{Lafferty},~\bfnm{John}\binits{J.}} \AND
  \bauthor{\bsnm{Wasserman},~\bfnm{Larry}\binits{L.}}
(\byear{2009}).
\btitle{The Nonparanormal: Semiparametric Estimation of High Dimensional
  Undirected Graphs}.
\bjournal{J. Mach. Learn. Research}
\bvolume{10}
\bpages{2295--2328}.
\end{barticle}
\endbibitem

\bibitem{MaathuisColombo14}
\begin{barticle}[author]
\bauthor{\bsnm{Maathuis},~\bfnm{Marloes~H.}\binits{M.~H.}} \AND
  \bauthor{\bsnm{Colombo},~\bfnm{Diego}\binits{D.}}
(\byear{2015}).
\btitle{A generalized back-door criterion}.
\bjournal{Ann. Statist.}
\bpages{1060-1088}.
\end{barticle}
\endbibitem

\bibitem{MaathuisColomboKalischBuehlmann10}
\begin{barticle}[author]
\bauthor{\bsnm{Maathuis},~\bfnm{M.~H.}\binits{M.~H.}},
  \bauthor{\bsnm{Colombo},~\bfnm{Diego}\binits{D.}},
  \bauthor{\bsnm{Kalisch},~\bfnm{Markus}\binits{M.}} \AND
  \bauthor{\bsnm{B\"uhlmann},~\bfnm{Peter}\binits{P.}}
(\byear{2010}).
\btitle{Predicting causal effects in large-scale systems from observational
  data}.
\bjournal{Nature Methods}
\bvolume{7}
\bpages{247--248}.
\end{barticle}
\endbibitem

\bibitem{MaathuisKalischBuehlmann09}
\begin{barticle}[author]
\bauthor{\bsnm{Maathuis},~\bfnm{M.~H.}\binits{M.~H.}},
  \bauthor{\bsnm{Kalisch},~\bfnm{M.}\binits{M.}} \AND
  \bauthor{\bsnm{B\"uhlmann},~\bfnm{P.}\binits{P.}}
(\byear{2009}).
\btitle{Estimating high-dimensional intervention effects from observational
  data}.
\bjournal{Ann. Statist.}
\bvolume{37}
\bpages{3133-3164}.
\end{barticle}
\endbibitem

\bibitem{MarbachEtAl10}
\begin{barticle}[author]
\bauthor{\bsnm{Marbach},~\bfnm{Daniel}\binits{D.}},
  \bauthor{\bsnm{Prill},~\bfnm{Robert~J.}\binits{R.~J.}},
  \bauthor{\bsnm{Schaffter},~\bfnm{Thomas}\binits{T.}},
  \bauthor{\bsnm{Mattiussi},~\bfnm{Claudio}\binits{C.}},
  \bauthor{\bsnm{Floreano},~\bfnm{Dario}\binits{D.}} \AND
  \bauthor{\bsnm{Stolovitzky},~\bfnm{Gustavo}\binits{G.}}
(\byear{2010}).
\btitle{Revealing strengths and weaknesses of methods for gene network
  inference}.
\bjournal{{PNAS}}
\bvolume{107}
\bpages{6286--6291}.
\end{barticle}
\endbibitem

\bibitem{Meek95}
\begin{binproceedings}[author]
\bauthor{\bsnm{Meek},~\bfnm{Christopher}\binits{C.}}
(\byear{1995}).
\btitle{Causal inference and causal explanation with background knowledge}.
In \bbooktitle{UAI 1995}.
\end{binproceedings}
\endbibitem

\bibitem{MooijHeskes13}
\begin{binproceedings}[author]
\bauthor{\bsnm{Mooij},~\bfnm{J.~M.}\binits{J.~M.}} \AND
  \bauthor{\bsnm{Heskes},~\bfnm{T.}\binits{T.}}
(\byear{2013}).
\btitle{Cyclic causal discovery from continuous equilibrium data}.
In \bbooktitle{UAI 2013}.
\end{binproceedings}
\endbibitem

\bibitem{NandyMaathuisRichardson14b}
\begin{bunpublished}[author]
\bauthor{\bsnm{Nandy},~\bfnm{P.}\binits{P.}},
  \bauthor{\bsnm{Maathuis},~\bfnm{M.~H.}\binits{M.~H.}} \AND
  \bauthor{\bsnm{Richardson},~\bfnm{T.~S.}\binits{T.~S.}}
(\byear{2016}).
\btitle{Supplement to ``{E}stimating the effect of joint interventions from
  observational data in sparse high-dimensional settings"}.
\end{bunpublished}
\endbibitem

\bibitem{Pearl09}
\begin{barticle}[author]
\bauthor{\bsnm{Pearl},~\bfnm{J.}\binits{J.}}
(\byear{2009}).
\btitle{Causal inference in statistics: An overview}.
\bjournal{Stat. Surv.s}
\bvolume{3}
\bpages{96-146}.
\end{barticle}
\endbibitem

\bibitem{PerkovicEtAl15}
\begin{binproceedings}[author]
\bauthor{\bsnm{Perkovic},~\bfnm{E}\binits{E.}},
  \bauthor{\bsnm{Textor},~\bfnm{J}\binits{J.}},
  \bauthor{\bsnm{Kalisch},~\bfnm{M}\binits{M.}} \AND
  \bauthor{\bsnm{Maathuis},~\bfnm{M.~H.}\binits{M.~H.}}
(\byear{2015}).
\btitle{A complete adjustment criterion}.
In \bbooktitle{UAI 2015}.
\end{binproceedings}
\endbibitem

\bibitem{Pourahmadi99}
\begin{barticle}[author]
\bauthor{\bsnm{Pourahmadi},~\bfnm{M.}\binits{M.}}
(\byear{1999}).
\btitle{Joint mean-covariance models with applications to longitudinal data:
  Unconstrained parameterisation}.
\bjournal{Biometrika}
\bvolume{86}
\bpages{677--690}.
\end{barticle}
\endbibitem

\bibitem{Ramsey06}
\begin{btechreport}[author]
\bauthor{\bsnm{Ramsey},~\bfnm{Joe}\binits{J.}}
(\byear{2006}).
\btitle{A {PC}-style Markov blanket search for high dimensional datasets.}
\btype{Technical Report} No. \bnumber{177},
\bpublisher{Philosophy, Carnegie Mellon University}.
\end{btechreport}
\endbibitem

\bibitem{Robins86}
\begin{barticle}[author]
\bauthor{\bsnm{Robins},~\bfnm{James}\binits{J.}}
(\byear{1986}).
\btitle{A new approach to causal inference in mortality studies with a
  sustained exposure period-application to control of the healthy worker
  survivor effect}.
\bjournal{Mathematical Modelling}
\bvolume{7}
\bpages{1393--1512}.
\end{barticle}
\endbibitem

\bibitem{RobinsHernanBrumback00}
\begin{barticle}[author]
\bauthor{\bsnm{Robins},~\bfnm{James~M.}\binits{J.~M.}},
  \bauthor{\bsnm{Hernan},~\bfnm{Miguel~Angel}\binits{M.~A.}} \AND
  \bauthor{\bsnm{Brumback},~\bfnm{Babette}\binits{B.}}
(\byear{2000}).
\btitle{Marginal structural models and causal inference in epidemiology}.
\bjournal{Epidemiology}
\bvolume{11}
\bpages{550-560}.
\end{barticle}
\endbibitem

\bibitem{SchaffterEtAl11}
\begin{barticle}[author]
\bauthor{\bsnm{Schaffter},~\bfnm{Thomas}\binits{T.}},
  \bauthor{\bsnm{Marbach},~\bfnm{Daniel}\binits{D.}} \AND
  \bauthor{\bsnm{Floreano},~\bfnm{Dario}\binits{D.}}
(\byear{2011}).
\btitle{Gene{N}et{W}eaver: {I}n silico benchmark generation and performance
  profiling of network inference methods}.
\bjournal{Bioinform.}
\bvolume{27}
\bpages{2263--2270}.
\end{barticle}
\endbibitem

\bibitem{ShpitserVanderWeeleRobins10}
\begin{binproceedings}[author]
\bauthor{\bsnm{Shpitser},~\bfnm{I.}\binits{I.}},
  \bauthor{\bsnm{VanderWeele},~\bfnm{T.~J.}\binits{T.~J.}} \AND
  \bauthor{\bsnm{Robins},~\bfnm{J.~M.}\binits{J.~M.}}
(\byear{2010}).
\btitle{On the validity of covariate adjustment for estimating causal effects}.
In \bbooktitle{UAI 2010}.
\end{binproceedings}
\endbibitem

\bibitem{SpirtesEtAl00}
\begin{bbook}[author]
\bauthor{\bsnm{Spirtes},~\bfnm{P.}\binits{P.}},
  \bauthor{\bsnm{Glymour},~\bfnm{C.}\binits{C.}} \AND
  \bauthor{\bsnm{Scheines},~\bfnm{R.}\binits{R.}}
(\byear{2000}).
\btitle{Causation, {P}rediction, and {S}earch},
\bedition{second} ed.
\bseries{Adaptive Computation and Machine Learning}.
\bpublisher{MIT Press}, \baddress{Cambridge}.
\end{bbook}
\endbibitem

\bibitem{StekhovenEtAl12}
\begin{barticle}[author]
\bauthor{\bsnm{Stekhoven},~\bfnm{Daniel~J.}\binits{D.~J.}},
  \bauthor{\bsnm{Moraes},~\bfnm{Izabel}\binits{I.}},
  \bauthor{\bsnm{Sveinbj\"{o}rnsson},~\bfnm{Gardar}\binits{G.}},
  \bauthor{\bsnm{Henning},~\bfnm{Lars}\binits{L.}},
  \bauthor{\bsnm{Maathuis},~\bfnm{Marloes~H.}\binits{M.~H.}} \AND
  \bauthor{\bsnm{B\"{u}hlmann},~\bfnm{Peter}\binits{P.}}
(\byear{2012}).
\btitle{Causal stability ranking}.
\bjournal{Bioinform.}
\bvolume{28}
\bpages{2819-2823}.
\end{barticle}
\endbibitem

\bibitem{TeramotoEtAl14}
\begin{barticle}[author]
\bauthor{\bsnm{Teramoto},~\bfnm{Reiji}\binits{R.}},
  \bauthor{\bsnm{Saito},~\bfnm{Chiaki}\binits{C.}} \AND
  \bauthor{\bsnm{Funahashi},~\bfnm{Shin-ichi}\binits{S.-i.}}
(\byear{2014}).
\btitle{Estimating causal effects with a non-paranormal method for the design
  of efficient intervention experiments}.
\bjournal{BMC Bioinform.}
\bvolume{15}
\bpages{1--14}.
\end{barticle}
\endbibitem

\bibitem{Tian01}
\begin{binproceedings}[author]
\bauthor{\bsnm{Tian},~\bfnm{Jin}\binits{J.}} \AND
  \bauthor{\bsnm{Pearl},~\bfnm{Judea}\binits{J.}}
(\byear{2001}).
\btitle{Causal discovery from changes}.
In \bbooktitle{UAI 2001}.
\end{binproceedings}
\endbibitem

\bibitem{vanderZarderEtAl14}
\begin{binproceedings}[author]
\bauthor{\bparticle{van~der} \bsnm{Zander},~\bfnm{Benito}\binits{B.}},
  \bauthor{\bsnm{Liskiewicz},~\bfnm{Maciej}\binits{M.}} \AND
  \bauthor{\bsnm{Textor},~\bfnm{Johannes}\binits{J.}}
(\byear{2014}).
\btitle{Constructing separators and adjustment sets in ancestral graphs}.
In \bbooktitle{UAI 2014}.
\end{binproceedings}
\endbibitem

\bibitem{VelenichGore13}
\begin{barticle}[author]
\bauthor{\bsnm{Velenich},~\bfnm{Andrea}\binits{A.}} \AND
  \bauthor{\bsnm{Gore},~\bfnm{Jeff}\binits{J.}}
(\byear{2013}).
\btitle{{The strength of genetic interactions scales weakly with mutational
  effects}}.
\bjournal{Genome Biol.}
\bvolume{14}
\bpages{R76}.
\end{barticle}
\endbibitem

\bibitem{Wright21}
\begin{barticle}[author]
\bauthor{\bsnm{Wright},~\bfnm{Sewall}\binits{S.}}
(\byear{1921}).
\btitle{Correlation and causation}.
\bjournal{J. Agric. Res.}
\bvolume{20}
\bpages{557--585}.
\end{barticle}
\endbibitem

\end{thebibliography}
